\documentclass{article}
\usepackage[german, english]{babel}
\usepackage[a4paper, left=2.5cm, right=2.5cm, top=2.5cm, bottom=2.5cm]{geometry}
\usepackage{amssymb}\usepackage{amsmath}
\usepackage{bbm}
\usepackage{amscd}
\usepackage{latexsym}
\usepackage{graphicx}
\usepackage{ifthen}
\usepackage{epsfig}
\usepackage{setspace}
\usepackage{subcaption}
\usepackage{hyperref}
\usepackage{dsfont}
\usepackage{mathrsfs}
\usepackage{color}
\usepackage{array}
\newcolumntype{P}[1]{>{\centering\arraybackslash}p{#1}}
\usepackage{makecell}
\usepackage{enumitem}
\usepackage[toc,page]{appendix}
\usepackage{float}
\hypersetup{
  colorlinks=true,
  linkcolor=blue,
  citecolor=magenta,
 }

\catcode`@=11
\renewcommand{\section}{\@startsection{section}{1}{0pt}{\medskipamount}
{\medskipamount}{\large\bf}}
\numberwithin{equation}{section}
\catcode`@=12

\newcommand{\R}{\mathds R}

\newcommand{\Acal}{{\cal A}}
\newcommand{\Fcal}{{\cal F}}
\newcommand{\Ecal}{{\cal E}}
\newcommand{\Bcal}{{\cal B}}

\newcommand{\Ical}{{\cal I}}

\newcommand{\ii}{\text{i}}
\newcommand{\ee}{\text{e}}
\newcommand{\dd}{\text{d}}

\newcommand{\Y}{Y_{j;m,n}}

\def\im{\mathrm{i}}

\def\pa{\mbox{$\partial$}}
\def\diff{\mathrm{d}}

\def\sfrac#1#2{{\textstyle\frac{#1}{#2}}}
\def\]{\right]}
\def\[{\left[}
\def\){\right)}
\def\({\left(}
\def\>{\rangle}
\def\<{\langle}
\def\+{\dagger}

\def\={\ =\ }
\def\und{\quad\textrm{and}\quad}
\def\with{\quad\textrm{with}\quad}
\def\for{\quad\textrm{for}\quad}

\begin{document}

\title{\bf\huge Trajectories of charged particles \\ in knotted electromagnetic fields}
\date{~}

\author{
{\Large Kaushlendra Kumar,}\ 
{\Large Olaf Lechtenfeld}\ and\ {\Large Gabriel Pican\c{c}o Costa}
\\[24pt]
Institut f\"ur Theoretische Physik and\\ 
Riemann Center for Geometry and Physics\\
Leibniz Universit\"at Hannover \\ 
Appelstra{\ss}e 2, 30167 Hannover, Germany
\\[24pt]
} 

\clearpage
\maketitle
\thispagestyle{empty}

\begin{abstract}
\noindent\large
We investigate the trajectories of point charges in the background of finite-action vacuum solutions of Maxwell's equations known as knot solutions. More specifically, we work with a basis of electromagnetic knots generated by the so-called `de Sitter method'. We find a variety of behaviors depending on the field configuration and the parameter set used. This includes an acceleration of particles by the electromagnetic field from rest to ultrarelativistic speeds, a quick convergence of their trajectories into a few narrow cones asymptotically for sufficiently high value of the coupling, and a pronounced twisting and turning of trajectories in a coherent fashion. This work is part of an effort to improve the understanding of knotted electromagnetic fields and the trajectories of charged particles they generate, and may be relevant for experimental applications.
\end{abstract}

	%
	%

\newpage
\setcounter{page}{1}

\section{Introduction}

\noindent
Electromagnetic knots were first developed in 1989 by Rañada \cite{Ranada89} using the Hopf map. These knots are finite-action vacuum solutions of Maxwell's equations that consist of rational functions in the spacetime coordinates. In the original construction, two complex scalar fields $\phi$ and $\theta$ are used, and their level curves coincide with the electric and magnetic field lines. Those fields can be seen as maps from $S^3 \times \R$ to $S^2$, where $S^3$ here is the compactified three-dimensional space $\R^3 \cup \{\infty\}$ and $S^2$ is the compactified complex plane $\mathbb{C} \cup \{\infty\}$. The solutions are characterized by a topological invariant, the so-called Hopf index. Since then, other approaches were developed to construct electromagnetic knot solutions, for example using the twistor theory developed by Penrose, complex Euler potentials, or special conformal transformations; see \cite{Arrayas17} for a comprehensive review. Knotted electromagnetic fields might become important for future applications for their unique characteristics. Therefore, it is important to seek experimental settings to generate those fields and to study scenarios with them. Irvine and Bouwmeester \cite{Irvine08} discuss the generation of knotted fields using Laguerre--Gaussian beams and predict potential applications in atomic particle trapping, the manipulation of cold atomic ensembles, helicity injection for plasma confinement, and in the generation of soliton-like solutions in a nonlinear medium. Laser beams with knotted polarization singularities were recently used to produce some simple knotted field configurations including the figure-8 knot in the lab \cite{LSMetal18}.

Lately, a new method \cite{OlafZhilin18} has been developed for deriving a complete basis of electromagnetic knotted solutions to Maxwell's equations. This was achieved by utilizing the conformal invariance of four-dimensional Maxwell theory and a conformal equivalence of half of de Sitter space dS$_4$ to the future part of Minkowski space $\R^{1,3}$. More explicitly, one utilizes a manifest SO($4$)-covariant formalism on the spatial three-sphere slices of dS$_4$ to obtain analytic solutions of Maxwell's equations in terms of hyperspherical harmonics, which can easily be mapped onto Minkowski space with an explicit conformal map. This method also reproduces the aforementioned Hopf--Rañada (HR) knot as a simple case. Several features of the electromagnetic fields constructed via this new method have been explored, such as fall-off behavior, asymptotic energy flow, null solutions, and conserved helicity and conformal charges \cite{KL20,KaushalGabriel21}. To seek experimental applications, however, it is essential to elucidate the behavior of charged particles in the background of these fields.

The objective of this paper is to study the behavior of classical point charges in the knotted electromagnetic fields obtained via the `de Sitter method'. We first review the spacetime correspondences used in the method, followed by the construction of the field configurations and a discussion of their properties with the help of illustrative figures of field lines and energy densities in Section \ref{sec2:fields}. Afterwards, in Section \ref{sec3:trajectories} we numerically solve the Lorentz force equation for relativistic classical charged particles subject to these fields in different settings and try to unravel the impact of various parameters on the trajectories of the particles.

\section{The construction of knotted electromagnetic fields}\label{sec2:fields}

\subsection{The ``de Sitter method"}

Four-dimensional de Sitter space $\diff{S}_4$ can be described as a hypersurface embedded in $\R^{1,4}$ and defined by the constraint
\begin{equation}\label{hyperbol}
	-q_0^2+q_1^2+q_2^2+q_3^2+q_4^2 \= \ell^2\ ,
\end{equation}
that is, a single-sheeted hyperboloid in $\R^{1,4}$, where the $q$'s are standard coordinates in the 5-dimensional Minkowski space, and $\ell$ is the so-called `de Sitter radius'. We can parametrize dS$_4$ using 
\begin{equation}
	q_0 \= -\ell \cot \tau \und q_{_A} \= \frac{\ell}{\sin \tau} \omega_{_A} \for A = 1,2,3,4\ ,
\end{equation}
with $\tau \in \Ical := (0,\pi)$ and $\omega_{_A}$ being coordinates of $\R^4$ embedding the unit three-sphere $S^3$ via $\omega_{_A} \omega_{_A} = 1$. The standard Minkowski metric on $\R^{1,4}$ then induces on d$S_4$ the metric
\begin{equation}
	\diff{s}^2 \= \frac{\ell^2}{\sin^2 \tau} \big( -\dd\tau^2 + \dd\Omega^2_3 \big)\ ,
\end{equation}
where $\dd\Omega^2_3$ is the round metric on the unit $S^3$. It is then clear that de Sitter space $\diff{S}_4$ is conformally equivalent to a cylinder $\Ical {\times} S^3$. We proceed to map part of the cylinder to the future half of four-dimensional Minkowski space $\R^{1,3}$ with
\begin{equation}\label{eq:coordinates}
	-\cot \tau \= \frac{t^2-r^2-\ell^2}{2\ell t}, \quad \omega_1 \= \sigma \frac{x}{\ell}, \quad \omega_2 \= \sigma \frac{y}{\ell}, \quad \omega_3 \= \sigma \frac{z}{\ell}, \quad \omega_4 \= \sigma \frac{r^2-t^2-\ell^2}{2\ell^2}\ ,
\end{equation}
where $x, y, z \in \R, t \in \R_+, r^2 = x^2+y^2+z^2$ and
\begin{equation}\label{eq:gamma}
	\sigma \= \frac{2\ell^2}{\sqrt{2\ell^2t^2+(r^2-t^2+\ell^2)^2}}\ .
\end{equation}
We can glue together two copies of the cylinder by taking $\tau \in 2\Ical := (-\pi,\pi)$ to cover the entire Minkowski space. If one expresses the metric in the $(t,x,y,z)$ coordinates one obtains
\begin{equation}
	\text{ds}^2 \= \frac{\ell^2}{t^2}\left(-\diff{t}^2 + \diff{x}^2 + \diff{y}^2 + \diff{z}^2\right)\ ,
\end{equation}
which shows the conformal equivalence between part of $2\Ical {\times} S^3$ and $\R^{1,3}$. For more details, see \cite{OlafZhilin18}.

Now we proceed to the construction of the Maxwell solutions. The fact that Maxwell's equations (or, more generally, the Yang--Mills equations) are conformally invariant in four dimensions allows one to solve them in any other four-dimensional spacetime that is conformally related to the desired spacetime. In particular, here we will get a basis of solutions on the Minkowski space by solving the equations on the cylinder over the three-sphere. This lets us take advantage of a SO($4$)-covariant formalism. Moreover, since $S^3$ is the group manifold of SU$(2)$, one can also parametrize the spatial part of the cylinder by the group elements of SU($2$).

Using the Maurer--Cartan prescription one obtains three anholonomic one-forms $e^a(\omega)$, with $a \in \{1,2,3\}$ and $\omega$ representing the embedding coordinates of the three-sphere. They can be computed using the self-dual 't Hooft symbol $\eta^a_{_{\ BC}}$: 
\begin{equation}\label{eq:one-form}
    e^a \= -\eta^a_{_{\ BC}}\, \omega_{_B}\, \diff \omega_{_C} \with \eta^a_{\ bc} \= \varepsilon_{abc} \und \eta^a_{\ b4} \= -\eta^a_{\ 4b} \= \delta^a_b\ .
\end{equation}
These one-forms satisfy the Maurer--Cartan equations and diagonalize the three-sphere metric, i.e.,
\begin{equation}\label{eq:maurercartan}
	\diff{e}^{a} + \epsilon_{abc}\,e^b\wedge e^c \= 0 \qquad \text{and} \qquad e^{a}\,e^{a} \= \dd\Omega^2_3\ .
\end{equation}
Let $\diff{\tau}$ be the one-form associated with the temporal coordinate on the cylinder. One can then expand the gauge connection one-form $\Acal$ on the cylinder as
\begin{equation}
	\Acal(\tau,\omega) \= X_\tau(\tau,\omega)\,\diff{\tau} + X_a(\tau,\omega)\,e^a\ ,
\end{equation}
where $X_\tau$ and $X_a$ are real functions on the cylinder. Using the temporal gauge-fixing condition, $X_\tau=0$, this simplifies to
\begin{equation}
	\Acal(\tau,\omega) \= X_a(\tau,\omega)\,e^a\ .
\end{equation}

Now, as mentioned before, we take advantage of the fact that we are working in $S^3$ to employ a SO($4$)-covariant formalism. The universal covering group of SO($4$) is spin($4$), which is equivalent to SU($2$)$_L\times$SU($2$)$_R$, where $L$ and $R$ stand for left and right. We can then use this structure to decompose the spatial dependence of functions on the cylinder using the hyperspherical harmonics $\Y(\omega)$ (also called left-right harmonics). For an explicit construction using the three-sphere coordinates, see \cite{OlafZhilin18}. Take $I_a$ and $J_a$ to be the generators of the two (left and right, respectively) $\mathfrak{su}(2)$ subalgebras, with
\begin{equation} \label{LieAlgebra}
	[I_a, I_b] \= \im\, \epsilon_{abc} I_c\ , \quad [J_a, J_b] \= \im\, \epsilon_{abc} J_c\ \und [I_a, J_b] \= 0\ .
\end{equation}
Let us define the ladder operators 
\begin{equation}
	I_{\pm} = (I_1 \pm \ii I_2)/\sqrt{2} \qquad \text{and} \qquad J_{\pm} = (J_1 \pm \ii J_2)/\sqrt{2}\ ,
\end{equation}
    such that the action of the $\mathfrak{su}(2)$ generators on the hyperspherical harmonics is given by
\begin{equation}
\begin{aligned}\label{actionJ}
	&I_{\pm}\, \Y = \sqrt{(j\mp m)(j\pm m + 1)/2}\, Y_{j;m\pm 1,n}\ , \;\;\;  J_{\pm}\, \Y = \sqrt{(j\mp n)(j\pm n + 1)/2}\, Y_{j;m,n\pm 1}\ , \\
	&I_3\, \Y = m\,\Y\ , \quad J_3\, \Y = n\, \Y\ , \und I^2\, \Y = J^2 \, \Y = j(j{+}1)\,\Y\ ,
\end{aligned}
\end{equation}
with $I^2 := I_a I_a$ and $J^2 := J_a J_a$ being the Casimirs of the two $\mathfrak{su}(2)$ subalgebras. We note that the differential of any smooth function $f \in C^\infty(2\Ical{\times} S^3)$ on the cylinder can be expanded as
\begin{equation}
     \diff{f} \= \diff{\tau}\,\pa_\tau f - 2\,\im\,e^a\,J_a f\ ,
\end{equation}
where $J_a$ is viewed as a differential operator (see \cite{KL20}).

On top of the temporal gauge, we can further impose the Coulomb gauge condition
\begin{equation}
	J_a\, X_a(\tau,\omega) \= 0\ .
\end{equation}
Moreover, the Maxwell equations $\diff{*\Fcal}=0$, with $\Fcal:=\diff{\Acal}$ read, in this setting,
\begin{equation}
	-\tfrac{1}{4} \partial_\tau^2 X_a \= (J^2 {+} 1) X_a + \ii\, \epsilon_{abc} J_b X_c\ .
\end{equation}

One can then expand
\begin{equation}
	X_a(\tau,\omega) \= \sum_{j=0}^{\infty}\, \sum_{m,n=-j}^{j} Z_a^{j;m,n}(\tau)\, \Y(\omega)
\end{equation}
to transform the gauge-fixing condition and the Maxwell equations into matrix equations diagonal in $j$ and $m$. Using $X_\pm(\tau,\omega) := X_1(\tau,\omega)\, \pm\, \ii\, X_2(\tau,\omega)$, the matrix equations can be decoupled and easily solved to find a full basis of solutions to the system of equations,
\begin{align}\label{eq:solutionsX}
	&X_+^{(j;m,n)}(\tau,\omega) \= \sqrt{(j-n)(j-n+1)/2}\; \ee^{\pm 2(j+1)\ii \tau}\, Y_{j;m,n+1}(\omega)\ ,\nonumber\\
	&X_3^{(j;m,n)}(\tau,\omega) \= \sqrt{(j+1)^2-n^2}\; \ee^{\pm 2(j+1)\ii \tau}\, Y_{j;m,n}(\omega)\ ,\\
	&X_-^{(j;m,n)}(\tau,\omega) \= -\sqrt{(j+n)(j+n+1)/2}\; \ee^{\pm 2(j+1)\ii \tau}\, Y_{j;m,n-1}(\omega)\ ,\nonumber
\end{align}
where $j\geq 0$, $m$ ranges from $-j$ to $j$, $n$ ranges from $-(j{+}1)$ to $j{+}1$ and it is understood that $\Y$ vanishes for $|n|>j$.

Now one can proceed to find the electromagnetic fields using 
\begin{equation}\label{eq:fieldfromA}
	\Fcal \= \diff{\Acal} \= \partial_\tau\,\Acal_a \diff{\tau}\wedge e^{a} - \left(\im\,J_{[b}\Acal_{c]} + \Acal_a \epsilon_{abc}\right) e^b \wedge e^c\ ,
\end{equation}
and the electric and magnetic field on the cylinder will have components $\Ecal_a = \Fcal_{\tau a}$ and $\Bcal_a = \tfrac{1}{2}\epsilon_{abc}\Fcal_{bc}$, respectively.

\subsection{Electromagnetic knots in Minkowski space}

\begin{figure}[h!]
\centering
   \includegraphics[width = 5cm, height = 5cm]{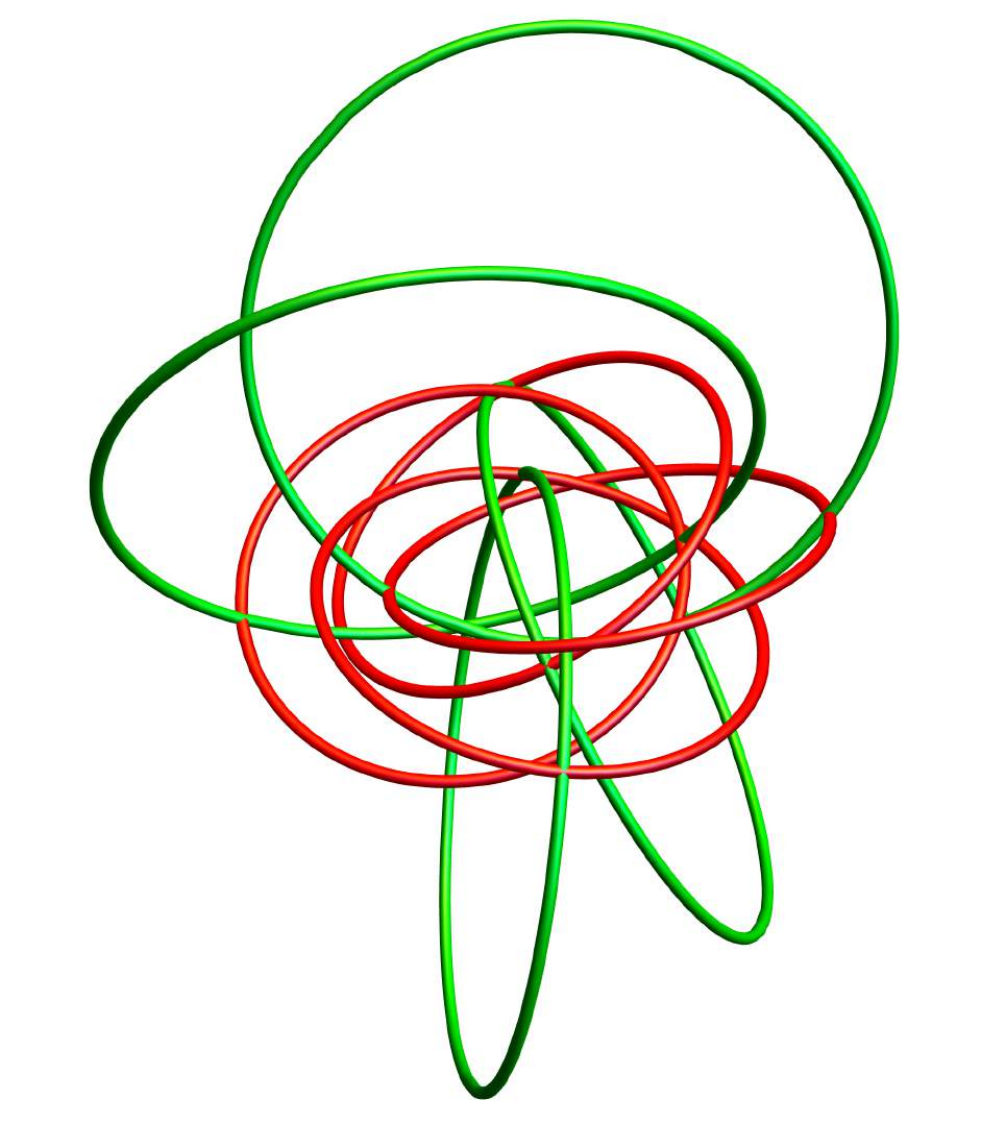}
   \includegraphics[width = 5cm, height = 5cm]{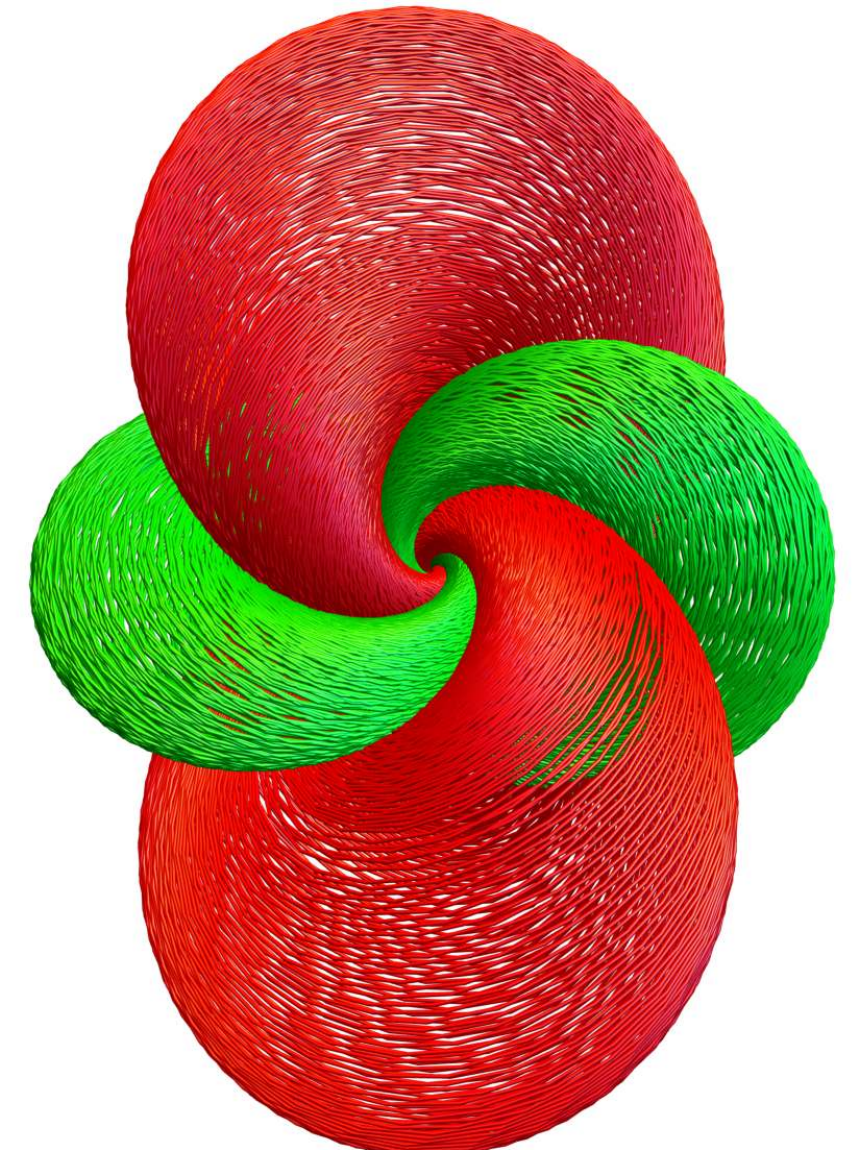}
   \includegraphics[width = 5cm, height = 5cm]{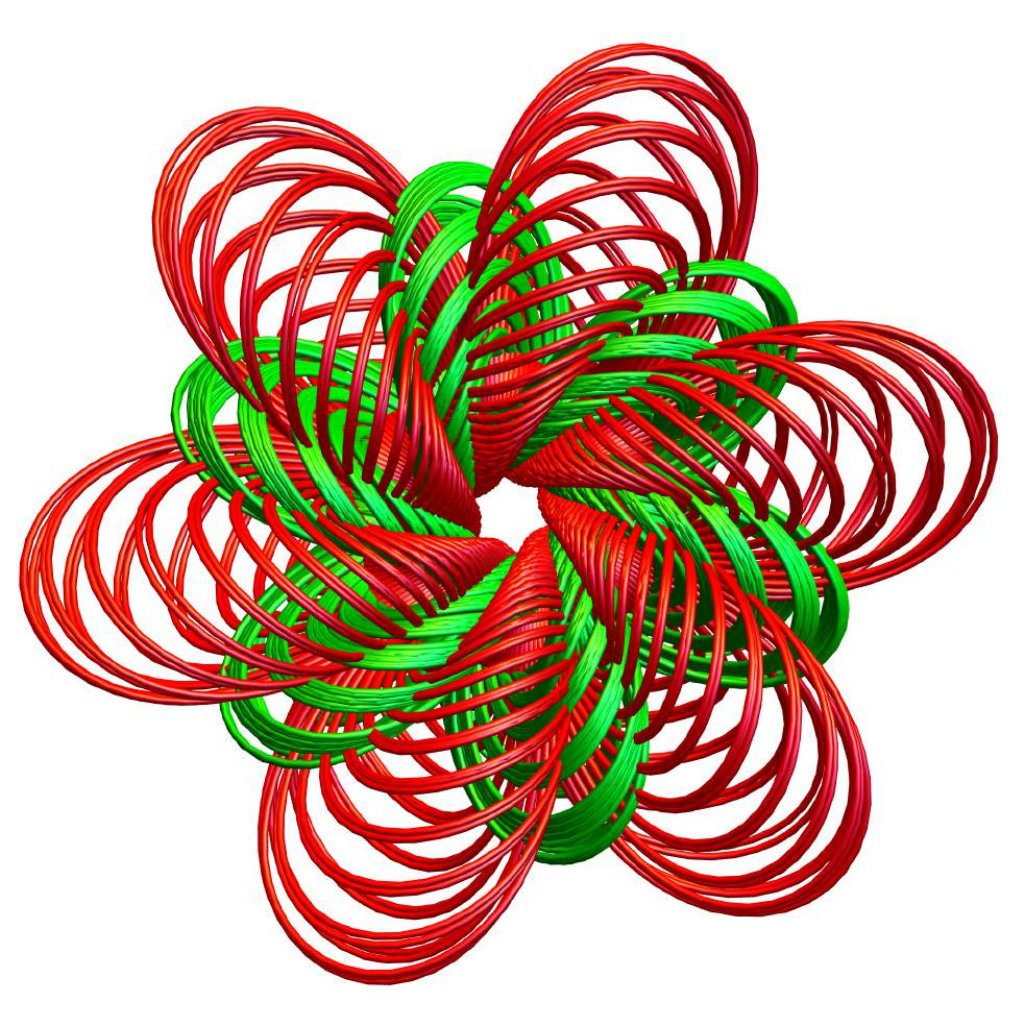}
 \caption{
 Electric (red) and magnetic (green) field lines at $t{=}0$ with 4 fixed seed points.
 Left: $(0;0,1)_I$ configuration. Center: $(\sfrac12;-\sfrac12,\sfrac32)_R$ configuration. Right: $(1;1,2)_I$ configuration. More self-knotted field lines start appearing with additional seed points in the simulation.}
\label{fieldLines}
\end{figure}

\begin{figure}[h!]
\centering
   \includegraphics[width = 5cm, height = 5cm]{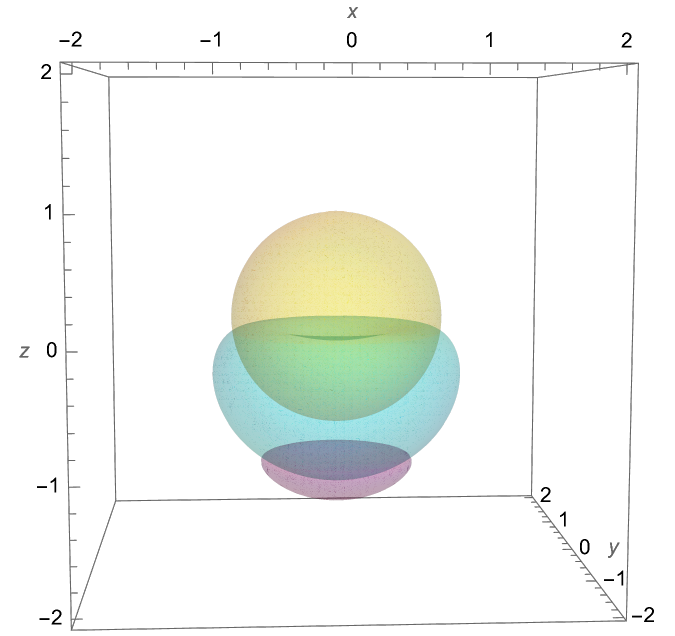}
   \includegraphics[width = 5cm, height = 5cm]{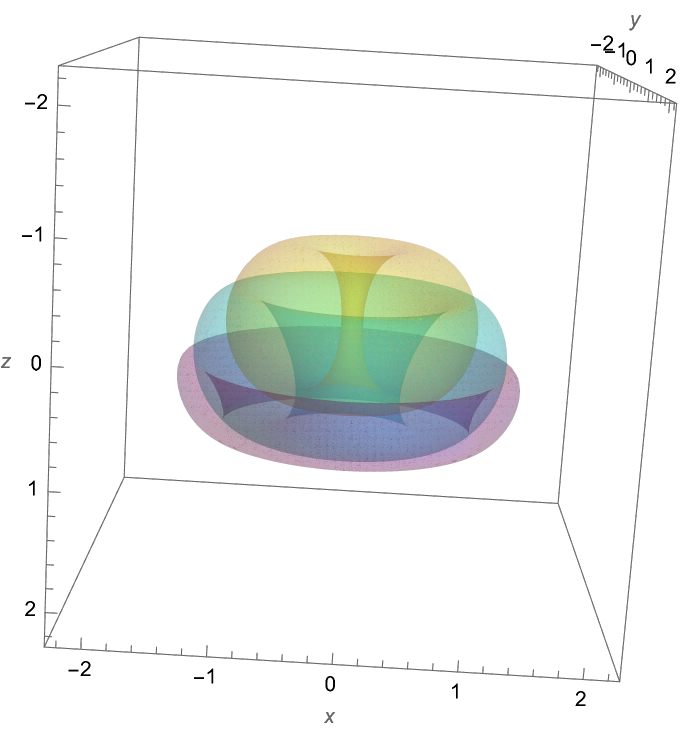}
   \includegraphics[width = 5cm, height = 5cm]{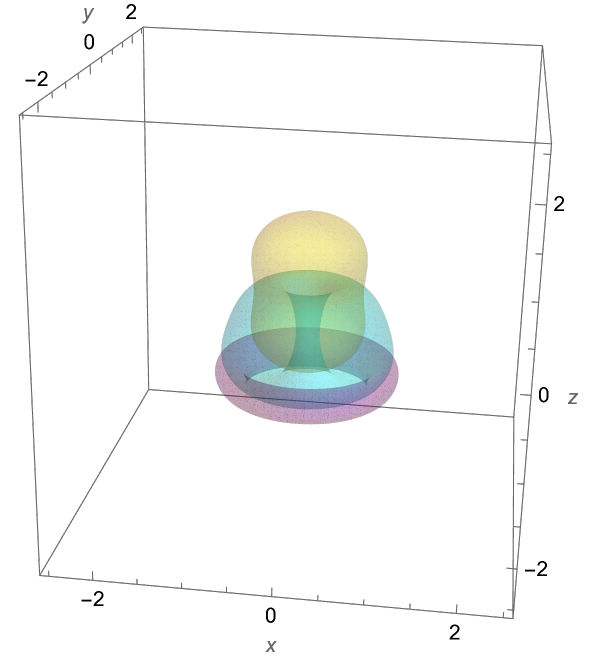}
 \caption{
 Contour plots for energy densities at $t{=}0$ (yellow), $t{=}1$ (cyan) and $t{=}1.5$ (purple) with contour value $0.9E_{max}(1.5)$.
 Left: $(0,0,1)_I$ configuration. Center: $(\sfrac12;-\sfrac12,-\sfrac32)_R$ configuration. Right: $(1;-1,1)_R$ configuration.}
\label{EnDen}
\end{figure}

To find the corresponding electromagnetic fields in Minkowski space, one has to write the one-forms $\dd \tau$ and $e^{a}$ in terms of $\dd x^\mu$. A straightforward computation using \eqref{eq:coordinates}, \eqref{eq:gamma} and \eqref{eq:one-form} gives these one-forms in terms of spacetime coordinates $(t,{\bf x})$:
\begin{align}
	\diff{\tau} &\= \frac{\sigma^2}{\ell^3} \left( \tfrac{1}{2}(t^2+r^2+\ell^2)\,\dd t - t\, x^k \dd x^k \right) \qquad \text{and}\\
	e^{a} &\= \frac{\sigma^2}{\ell^3} \left[ t\, x^{a} \dd t - \left(\tfrac{1}{2}(t^2-r^2+\ell^2) \delta^{ak} + x^{a} x^{k} + \ell\, \epsilon_{ajk} x^j \right) \dd x^k \right]\ .
\end{align}
Substituting into (\ref{eq:fieldfromA}), one gets $F_{\mu \nu}$ from $\Fcal = \tfrac{1}{2}F_{\mu\nu}\, \dd x^\mu\, \dd x^\nu$, with $\dd x^0 = \dd t$, and obtains the components $E_i = F_{0i}$ and $B_i = \tfrac{1}{2} \epsilon_{ijk}F_{jk}$ of the electric and magnetic fields in Minkowski space for any configuration generated by the basis (\ref{eq:solutionsX}). Since the basis configurations \eqref{eq:solutionsX} are complex, the corresponding fields on Minkowski space will also be complex. Hence, they combine two physical solutions, namely the real and imaginary parts, which we denote as
\begin{equation*}
    (j;m,n)_R~ \textrm{configuration} \und (j;m,n)_I ~\textrm{configuration}\ ,
\end{equation*}
respectively. The basis configurations increase in complexity with increasing $j$, as shown in Figure \ref{fieldLines}. Furthermore, these fields have a preferred $z$-direction due to our convention to diagonalize the $J_3$ action in \eqref{actionJ} (notice here that the SO(3) isometry subgroup, and hence its generators $J_a$, are identified on the cylinder and the Minkowski side; see \cite{KL20} for details). This is clearly exemplified in Figure \ref{EnDen}, where the energy density $E := \sfrac12 ({\bf E}^2 + {\bf B}^2)$ decreases along the $z$-axis. As a result, the basis fields along the $z$-axis (i.e.~${\bf E}(t,x{=}0,y{=}0,z)$ and ${\bf B}(t,x{=}0,y{=}0,z)$) are either directed in the $xy$-plane or along the $z$-axis. In fact, for extreme field configurations $(j;\pm j,\pm (j{+}1))$, for any $j{>}0$, the fields along the $z$-axis vanish for all times. In the simulations we have also used the maximum of the energy density at time $t$, i.e.~$E_{max}(t)$ (that occurs at several points ${\bf x}_{max}$ that are located symmetrically with respect to the origin), for different initial conditions and field configurations, and in each case we have employed a parameter $R_{max}(t)$ of `maximal' radius defined via
\begin{equation}\label{Rmax}
    E\left( t,{\bf x}_{max}(t) \right) \= E_{max}(t)\quad \implies \quad R_{max}(t)\ :=\ |{\bf x}_{max}(t)|\ .
\end{equation}

As previously stated, the celebrated HR knot \cite{OlafZhilin18} turns out to be the same as our $(0;0,1)_I$ basis configuration. One can also construct generalizations of HR-knots such as the time-translated and rotated Hopfions using a linear combination of $j{=}0$ configurations \cite{KaushalGabriel21}. Moreover, we find that some of our basis configurations are related to the (p,q)-torus knots arising from Bateman's construction \cite{Arrayas17}. We illustrate this point in Figure \ref{fieldLines} where we find the following correspondences:
\begin{equation}
	\textrm{Hopfian} \leftrightarrow (0,0,1)_I \leftrightarrow (1,1)\ ,\qquad (\sfrac12,-\sfrac12,\sfrac32)_R \leftrightarrow (2,1)\ ,\qquad (1,1,2)_I \leftrightarrow (1,3)\ . 
\end{equation}


\section{Trajectories}\label{sec3:trajectories}

Given a knotted electromagnetic field configuration, a natural issue that arises is the behavior of charged particles in the background of such a field. This question was already discussed in the context of the Hopfion \cite{Arrayas10}. Now we proceed to address this issue in the context of the `de Sitter' method by analyzing, with numerical simulations (see Mathematica Notebook \cite{nbFile}), the trajectories of several (identical) charged point particles for the family of knotted field configurations that we encountered in the last section. We will consider basis field configurations (up to $j{=}1$) here for simplicity.

The trajectories of these particles are governed by the relativistic Lorentz equation
\begin{equation}\label{eq:force1}
	\frac{\dd {\bf p}}{\dd t} = q({\bf E}_{\ell}+{\bf v} \times {\bf B}_{\ell})\ ,
\end{equation}
where $q$ is the charge of the particle, ${\bf p} {=} \gamma m {\bf v}$ is the relativistic three-momentum, ${\bf v}$ is the usual three-velocity of the particle, $m$ is its mass, $\gamma {=} (1-{\bf v}^{\,2})^{-1/2}$ is the Lorentz factor, and ${\bf E}_{\ell}$ and ${\bf B}_{\ell}$ are dimensionful electric and magnetic fields respectively. With the energy of the particle $E_{p} {=} \gamma m$ and $\dd E_p/ \dd t = q\, {\bf v} \cdot {\bf E}$, one can rewrite (\ref{eq:force1}) in terms of the derivative of ${\bf v}$ \cite{landau2} as
\begin{equation}\label{eq:force2}
	\frac{\dd {\bf v}}{\dd t} = \frac{q}{\gamma m} \left( {\bf E}_{\ell} + {\bf v} \times {\bf B}_{\ell} - ({\bf v} \cdot {\bf E}_{\ell})\,{\bf v} \right)\ .
\end{equation}
Equations (\ref{eq:force1}) and (\ref{eq:force2}) are equivalent, and either one can be used for a simulation purpose; they only differ by the position of the nonlinearity in ${\bf v}$. In natural units $\hbar {=} c {=} \epsilon_0 {=} 1$, every dimensionful quantity can be written in terms of a length scale. We relate all dimensionful quantities to the de Sitter radius $\ell$ from equation (\ref{hyperbol}) and work with the corresponding dimensionless ones as follows:
\begin{equation}
    T := \frac{t}{\ell}\ ,\quad {\bf X} := \frac{{\bf x}}{\ell}\ , \quad {\bf V} := \frac{\diff X}{\diff T} \equiv {\bf v}\ , \quad {\bf E} := \ell^2 {\bf E}_{\ell}, \und {\bf B} := \ell^2 {\bf B}_{\ell}\ .
\end{equation}
Moreover, the fields are solutions of the homogeneous (source-free) Maxwell equations, so they can be freely rescaled by any dimensionless constant factor $\lambda$. Combining the above considerations, one can rewrite (\ref{eq:force1}) (o    r analogously for (\ref{eq:force2})) fully in terms of dimensionless quantities as
\begin{equation}
		\frac{\dd (\gamma{\bf V})}{\dd T} = \kappa({\bf E} + {\bf V} \times {\bf B})\ ,
\end{equation}
where $\kappa = \sfrac{q\ell^3\lambda}{m}$ is a dimensionless parameter. One consequence of this parameter is that we can tune the values of each of the constants separately. In particular, we can make the charge as small as needed without changing $\kappa$ such that the effect of the backreaction on the trajectories becomes negligible. As for the initial conditions, we mostly work in the following two main scenarios:
\begin{enumerate}[label=(\arabic*)]
\item $N$ identical charged particles with ${\bf V}_0 {\equiv} {\bf V}(T{=}0) {=} 0$ located symmetrically (with respect to the origin), or

\item $N$ identical charged particles with $\,{\bf X}_0 {\equiv} {\bf X}(T{=}0) {=} 0\,$ with particle velocities directed radially outward in a symmetric fashion (with respect to the origin; shown in colored arrows),
\end{enumerate}
with the following 3 sub-cases for both of these conditions:
\begin{enumerate}[label=(\Alph*)]
    \item Along a line,
    \item On a circle of radius $r$,
    \item On a sphere of radius $r$.
\end{enumerate}

\begin{figure}[H]
\centering
   \includegraphics[width = 7.5cm, height = 5cm]{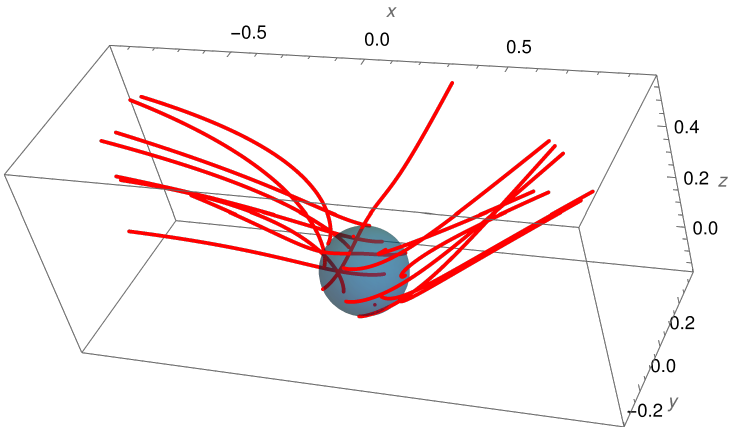}
   \includegraphics[width = 7.5cm, height = 5cm]{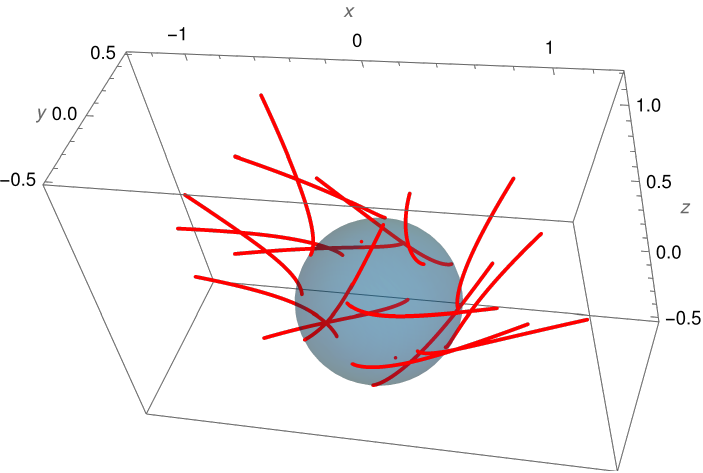}
 \caption{Simulation of $N{=}18$ particles in scenario (1C) for $(\sfrac12;-\sfrac12,-\sfrac32)_R$ with $\kappa{=}10$ and for $t\in[0,1]$. Left: $r{=}R_{max}(0){\approx}0.447$. Right: $r{=}R_{max}(0)/3$.}
\label{TrajRmaxneq0}
\end{figure}

We vary several parameters including the initial conditions with different directions of lines and planes for each configuration, the value of $\kappa$, and the simulation time in order to study the behavior of the trajectories. In several field configurations studied below, we find that $R_{max}(0)=0$, so we use a small radius $r$ for the initial condition of kind (1) to be able to probe the particles around a region of maximum energy of the field. In this scenario, the effect of the field on the trajectories of the particles is more prominent, as expected, and this helps us understand small perturbations of the trajectories as compared to a particle starting at rest from the origin. The effect of the fields on particles starting near the maximum of the energy density is also more prominent for $R_{max}(0){\neq}0$, as illustrated in Figure \ref{TrajRmaxneq0}. Moreover, for the initial condition of kind (2) we use the particle initial speeds in the range where it is (i) non-relativistic, (ii) relativistic (usually between $0.1$ and $0.9$), and (iii) ultrarelativistic (here, $0.99$ or higher).

\begin{figure}[H]
\centering
   \includegraphics[width = 5cm, height = 5cm]{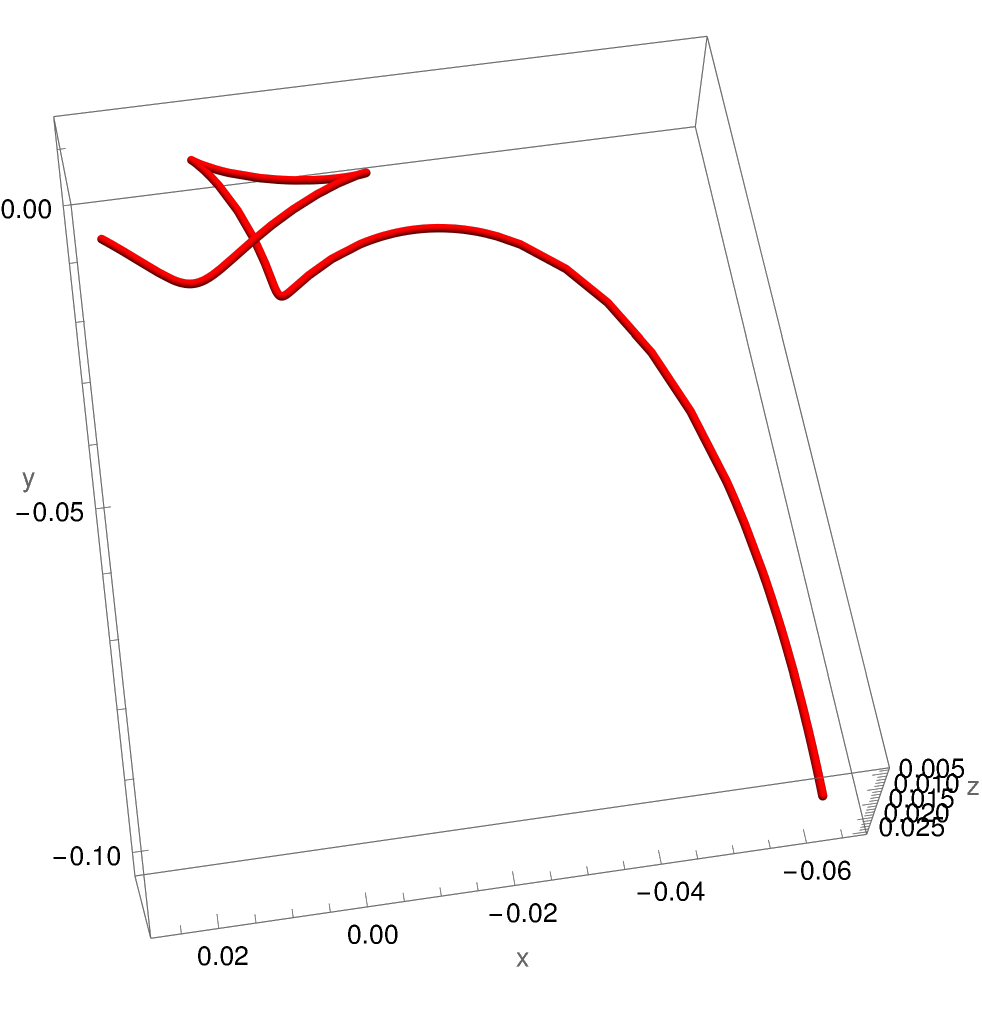}
   \includegraphics[width = 5cm, height = 5cm]{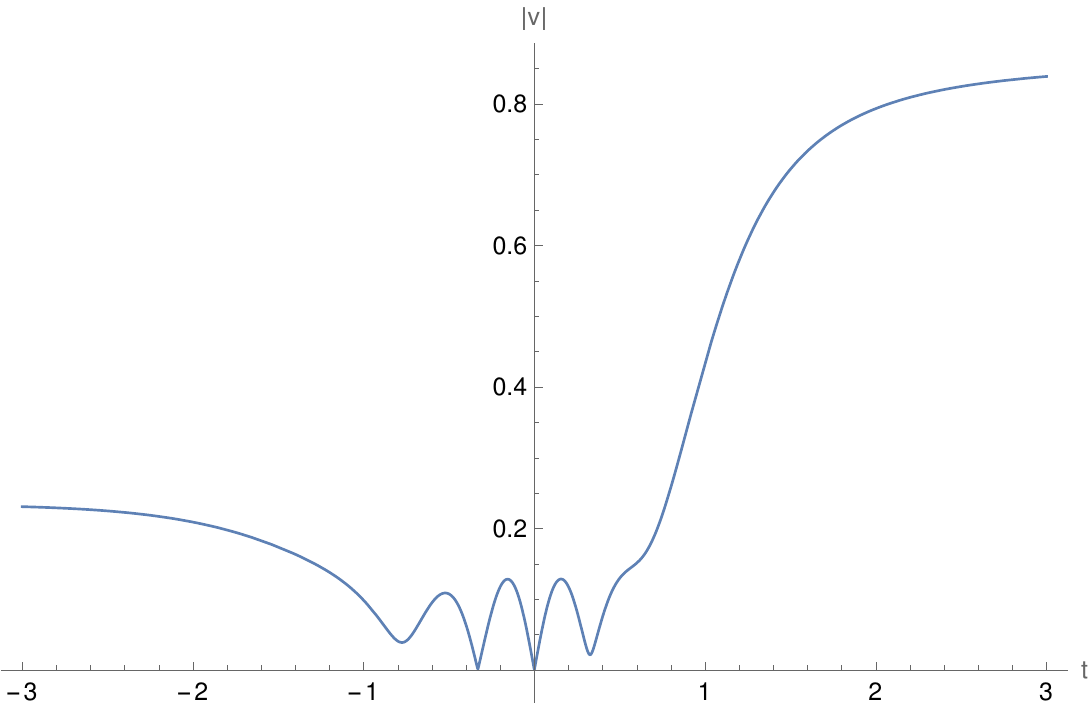}
   \includegraphics[width = 5cm, height = 5cm]{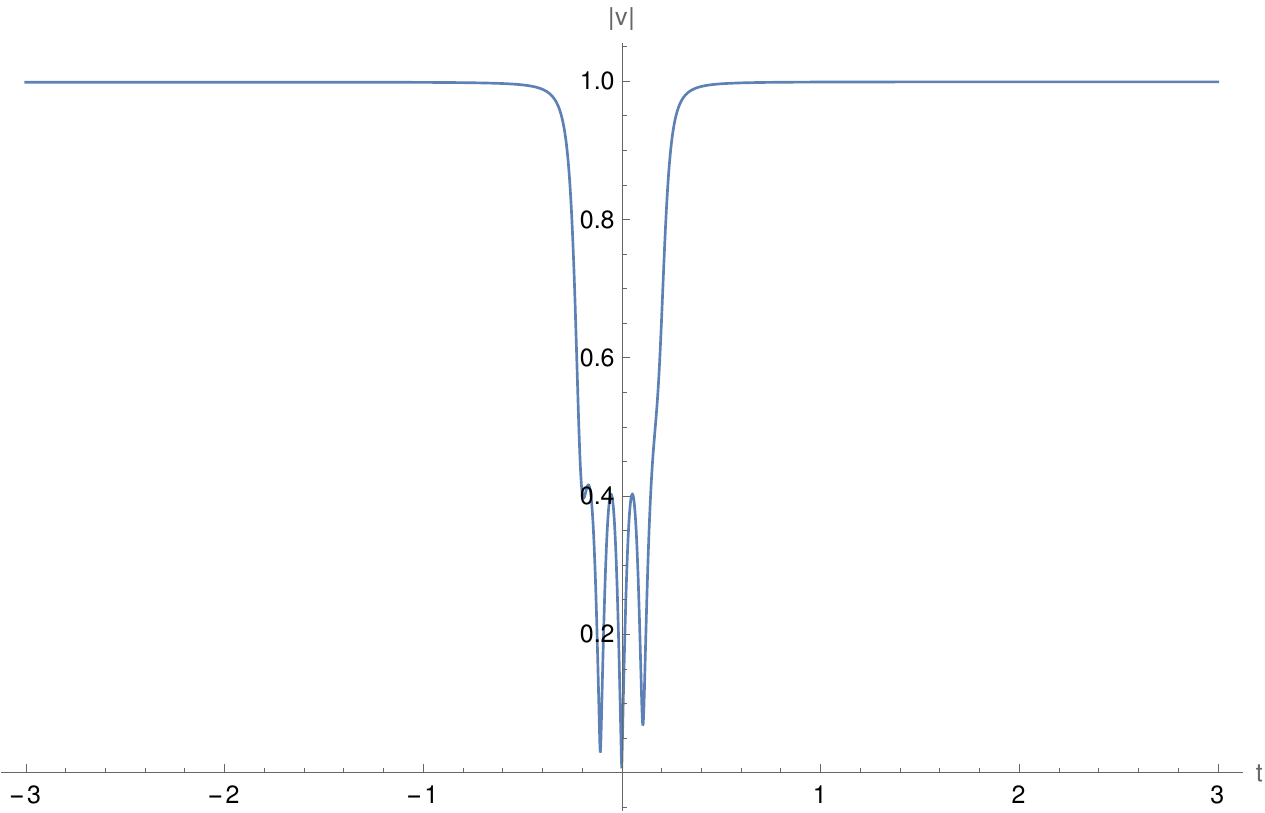}
 \caption{
 Trajectory of a charged particle for $(\sfrac12;-\sfrac12,-\sfrac32)_R$ configuration with initial conditions ${\bf X}_0 {=} (0.01,0.01,0.01)$ and ${\bf V}_0 {=} 0$ simulated for $t \in [-1,1]$.
 Left: Particle trajectory. Center: absolute velocity profile for $
\kappa {=} 10$. Right: absolute velocity profile for $\kappa {=} 100$.}
\label{singleTrajectory}
\end{figure}

We observe a variety of different behaviors for these trajectories, some of which we summarize below with the aid of figures. Firstly, it is worth noticing that, even with all fields decreasing as powers of both space and time coordinates, in most field configurations we observe particles getting accelerated from rest up to ultrarelativistic speeds. The limit of these ultrarelativistic speeds for higher times depend on the magnitude of the fields (see, for example, Figure \ref{singleTrajectory}).

\begin{figure}[H]
\centering
   \includegraphics[width = 7cm, height = 5cm]{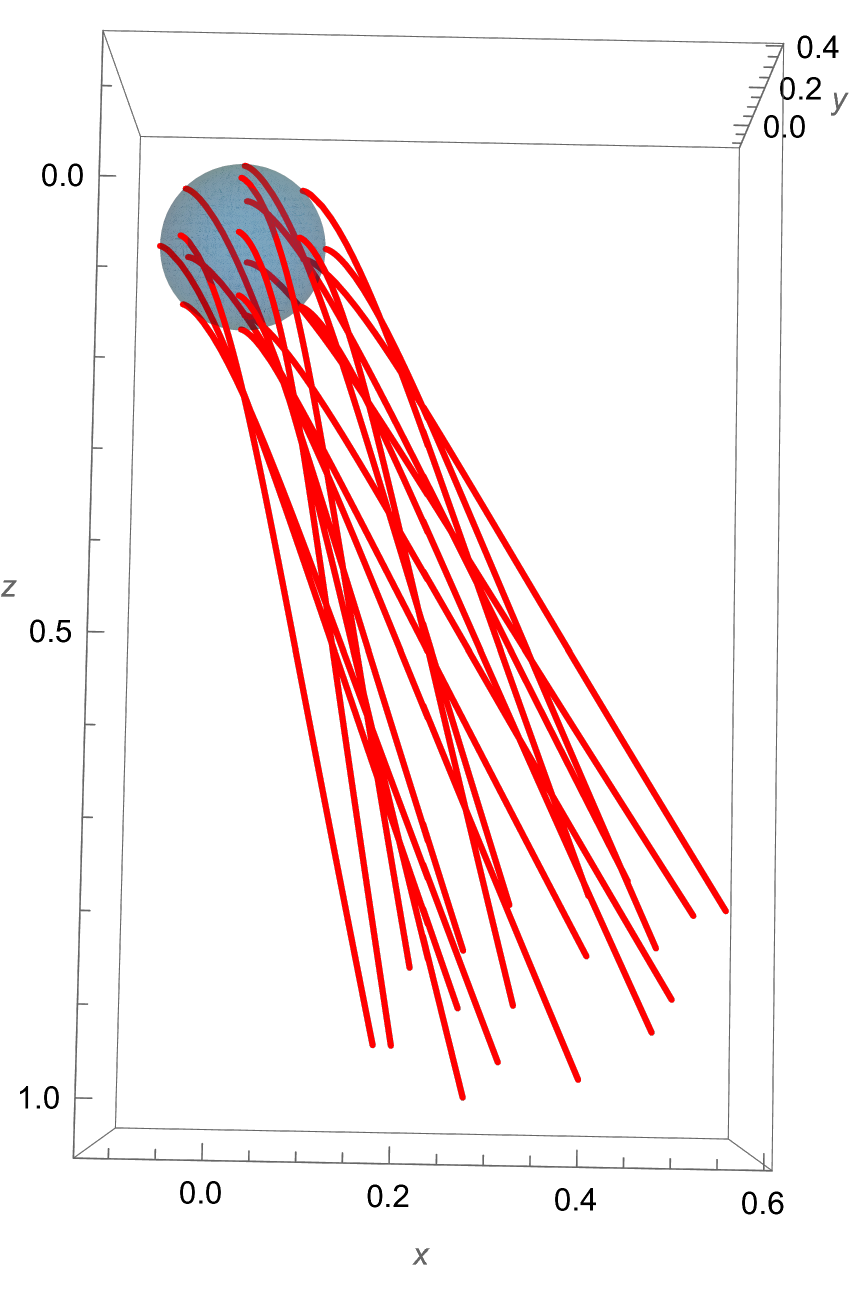}\hspace{1cm}
   \includegraphics[width = 7cm, height = 5cm]{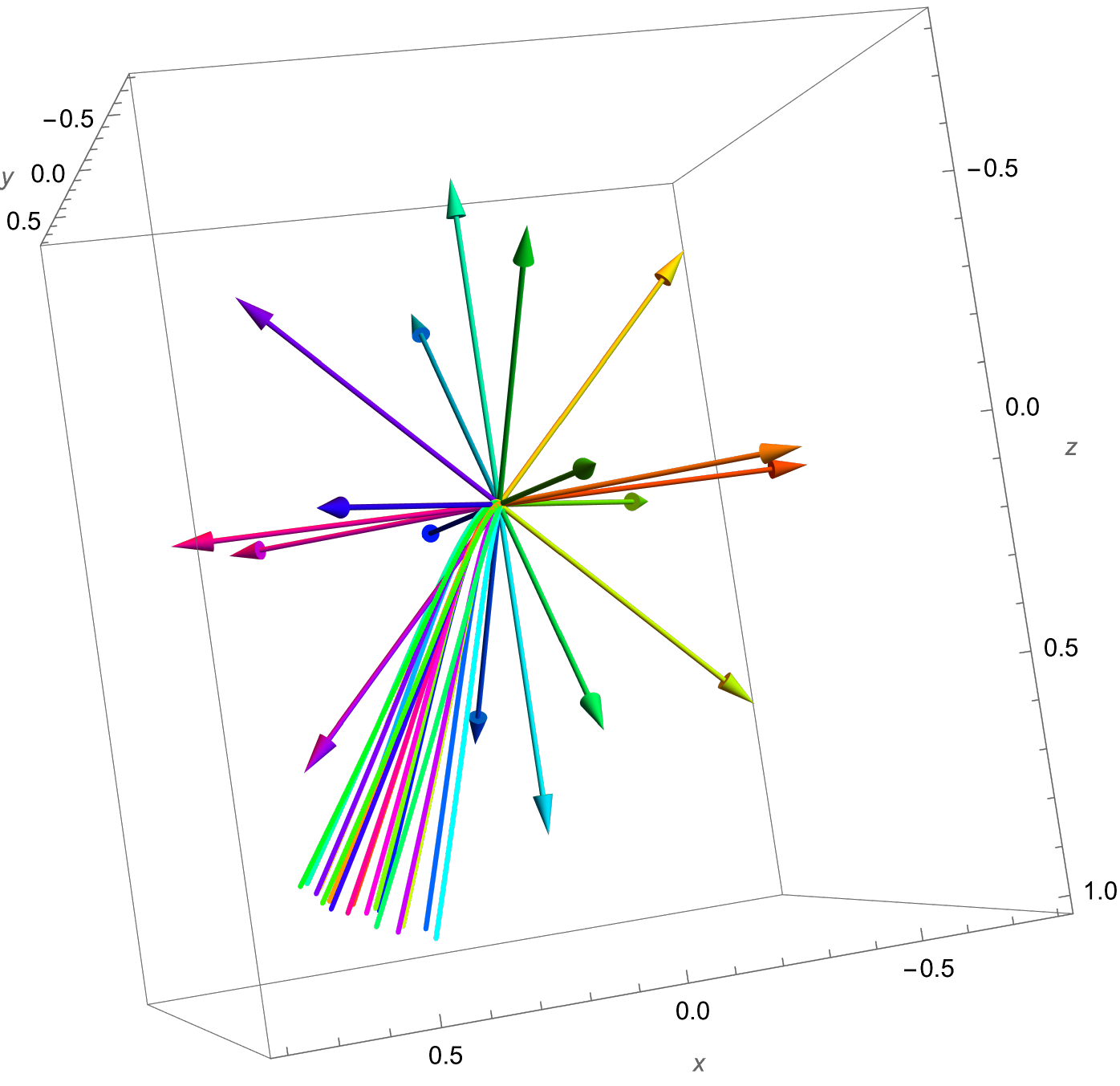}
 \caption{
 Simulation of $N{=}18$ particles for $(0,0,-1)_I$ configuration, with $\kappa {=} 100$ and $t\in[0,1]$. Left: scenario (1C) with $r{=}0.1$. Right: scenario (2C) with $r{=}0.75$.}
\label{j001plots}
\end{figure}

\begin{figure}[H]
\centering
   \includegraphics[width = 7cm, height = 5cm]{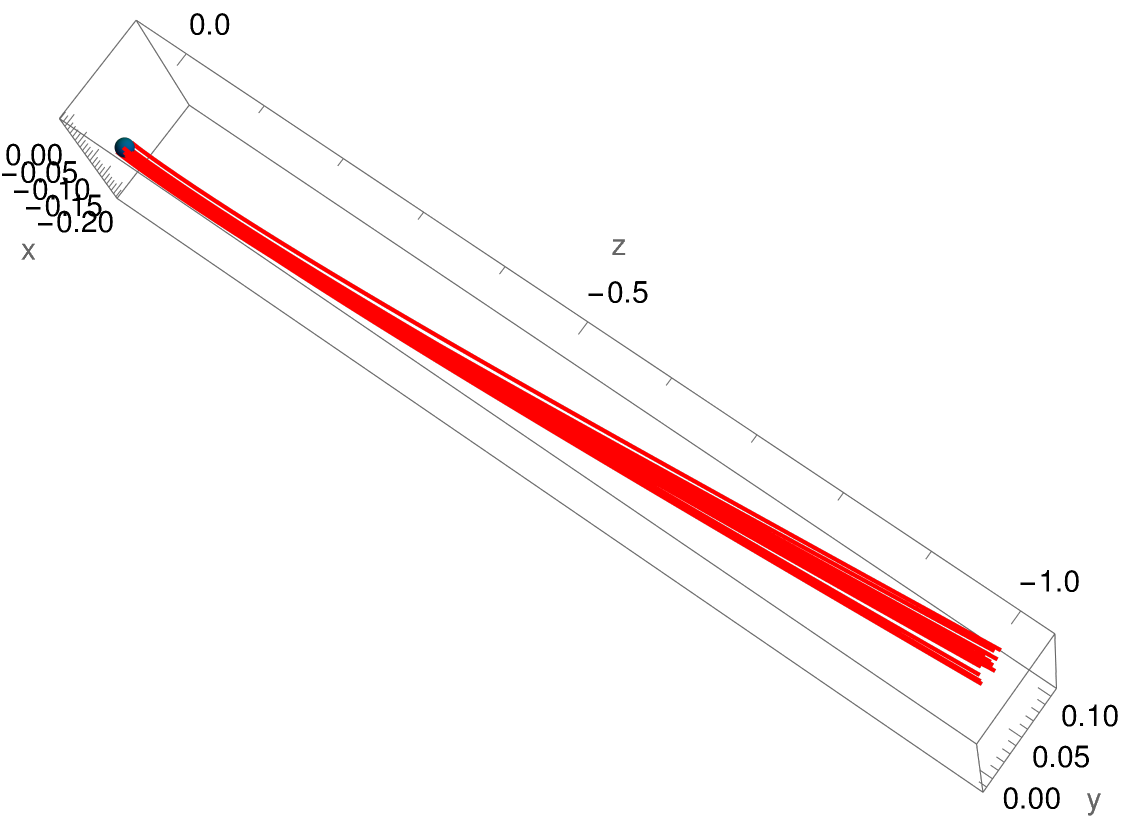}\hspace{1cm}
   \includegraphics[width = 7cm, height = 5cm]{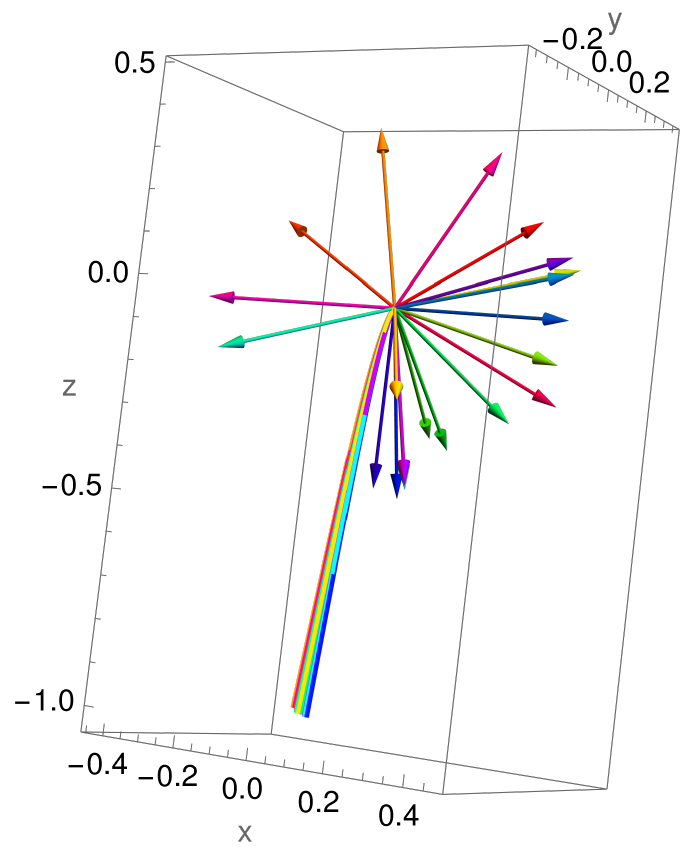}
 \caption{
    Simulation of $N{=}20$ particles for $(0,0,1)_I$ configuration, with $\kappa {=} 1000$ and $t\in[0,1]$. Left: particles starting from rest and located randomly inside a solid ball of radius $r{=}0.01$ ($R_{max}=0$). Right: particles located at origin and directed randomly (shown with colored arrows) with $|{\bf V}_0|{=}0.45$.}
\label{HRplots}
\end{figure}

With fixed initial conditions (of kind (1) or (2)) and for higher values of $\kappa$ one can expect, in general, that the initial conditions may become increasingly less relevant. For some fields configurations we indeed found that, with increasing $\kappa$, the particles get more focused and accumulate like a beam of charged particles along some specific region of space and move asymptotically for higher simulation times. This is exemplified below with two $j{=}0$ configurations: the $(0,0,-1)_I$ configuration in Figure \ref{j001plots}, and the HR configuration in Figure \ref{HRplots}. We have verified this feature not just with symmetric initial conditions of particles like that with initial conditions (1) and (2) (as in Figure \ref{j001plots}), but also in several initial conditions asymmetric with respect to the origin, like particles located randomly inside a sphere of fixed radius about the origin with zero initial velocity, and particles located at the origin but with different magnitudes of velocities. Figure \ref{HRplots} is an illustrative example for both of these latter scenarios of asymmetric initial conditions.

\begin{figure}[H]
\centering
   \includegraphics[width = 7cm, height = 5cm]{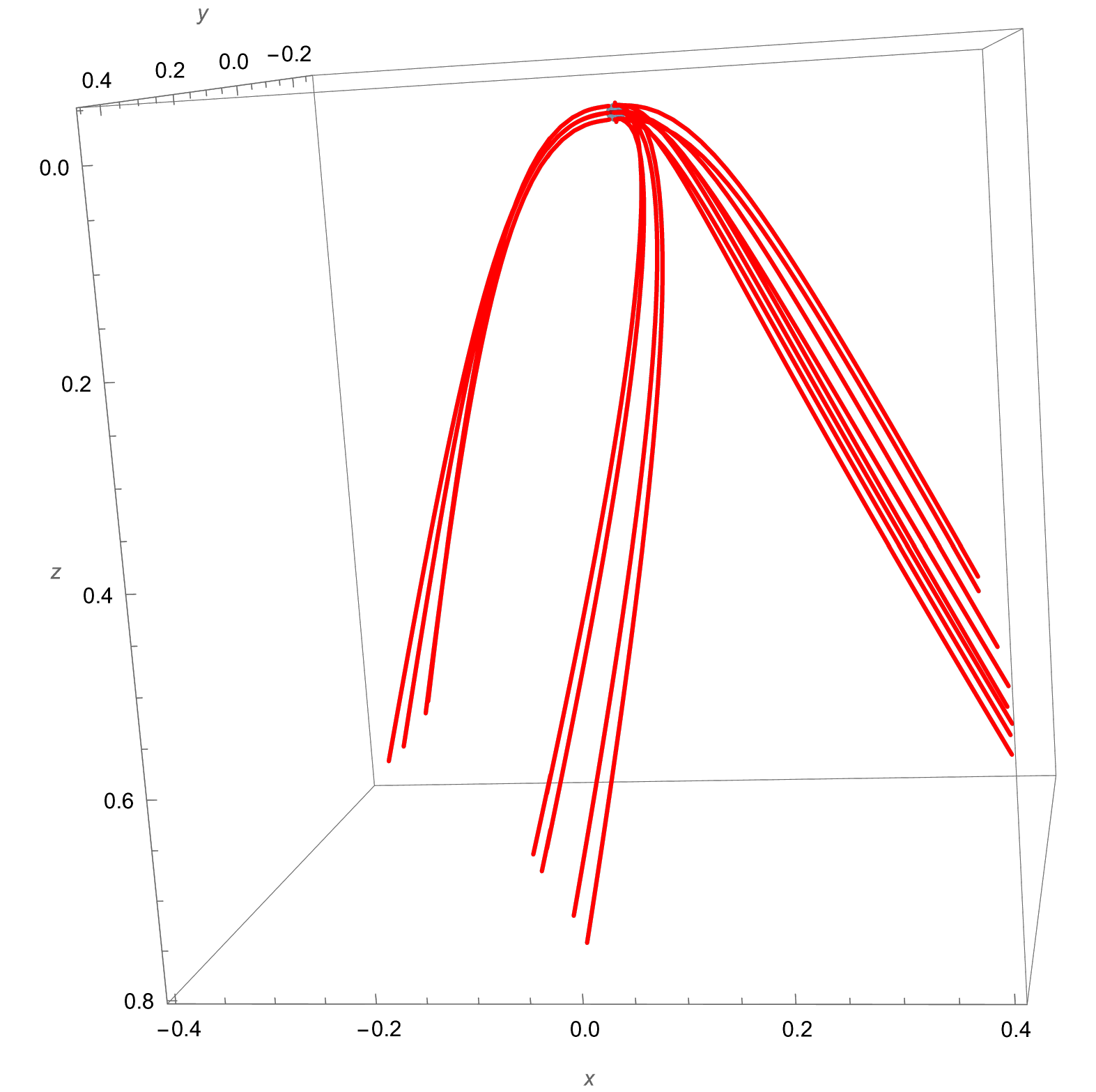}\hspace{1cm}
   \includegraphics[width = 7cm, height = 5cm]{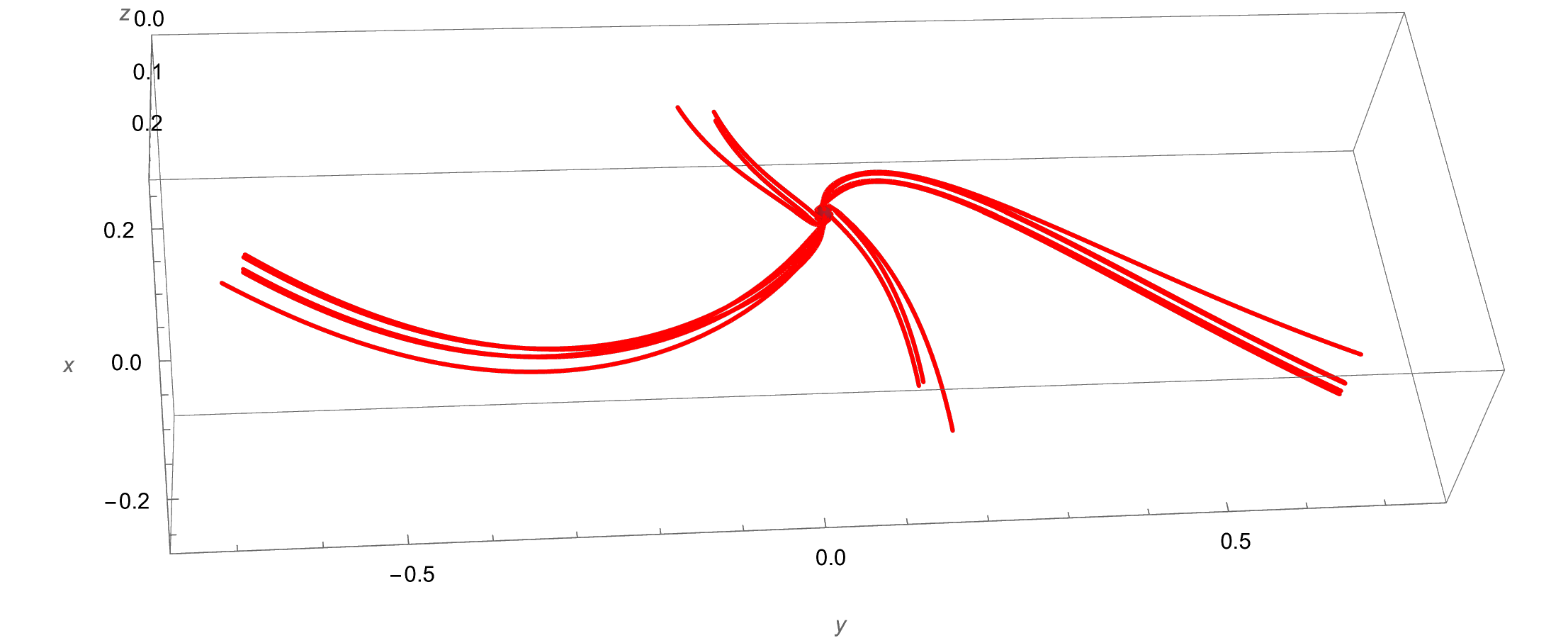}
 \caption{
 Simulation of $N{=}18$ particles in scenario (1C) with $r{=}0.01$ and for $t\in[0,3]$. Left: $(1,-1,-2)_R$ configuration with $\kappa {=} 500$. Right: $(1,-1,-1)_R$ configuration with $\kappa {=} 10$.}
\label{3-4furcation}
\end{figure}

This is not always the case though. For some $j{=}\tfrac12$ and $j{=}1$ configurations, and with initial particle positions in a sphere of very small radius about the origin, we are able to observe the splitting of particle trajectories (starting in some specific solid angle regions around the origin) into two, three or even four such asymptotic beams that converge along some particular regions of space (depending on the initial location of these particles in one of these solid angle regions). Trajectories generated by two such $j{=}1$ configurations have been illustrated in Figure \ref{3-4furcation}. 
%

\begin{figure}[H]
\centering
   \includegraphics[width = 7cm, height = 5cm]{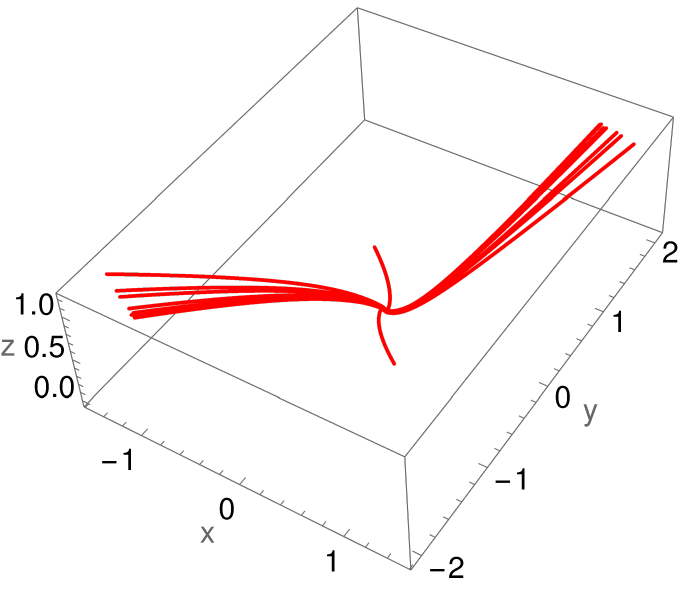}\hspace{1cm}
   \includegraphics[width = 7cm, height = 5cm]{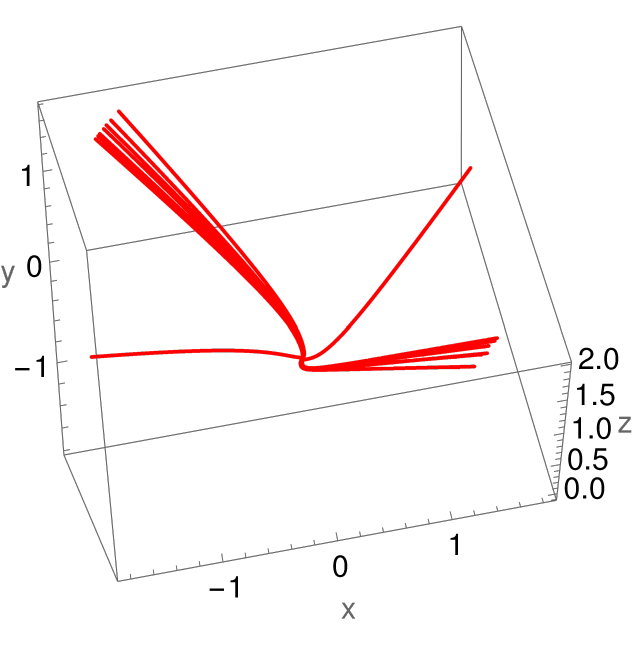}
 \caption{
 Simulation of $N{=}18$ particles in scenario (1C) with $\kappa{=}10$, $r{=}0.01$ and for $t\in[0,3]$. Left: $(\sfrac12;-\sfrac12,-\sfrac32)_R$ configuration. Right: $(1,0,-2)_I$ configuration.}
 \label{Bifurcation}
\end{figure}

Naturally, there are also regions of unstable trajectories for particles starting between these solid angle regions (see Figure \ref{Bifurcation}), which generally include the preferred $z$-axis, since in some cases trajectories that start at rest in the $z$-axis never leave it.

\begin{figure}[H]
\centering
   \includegraphics[width = 7cm, height = 5cm]{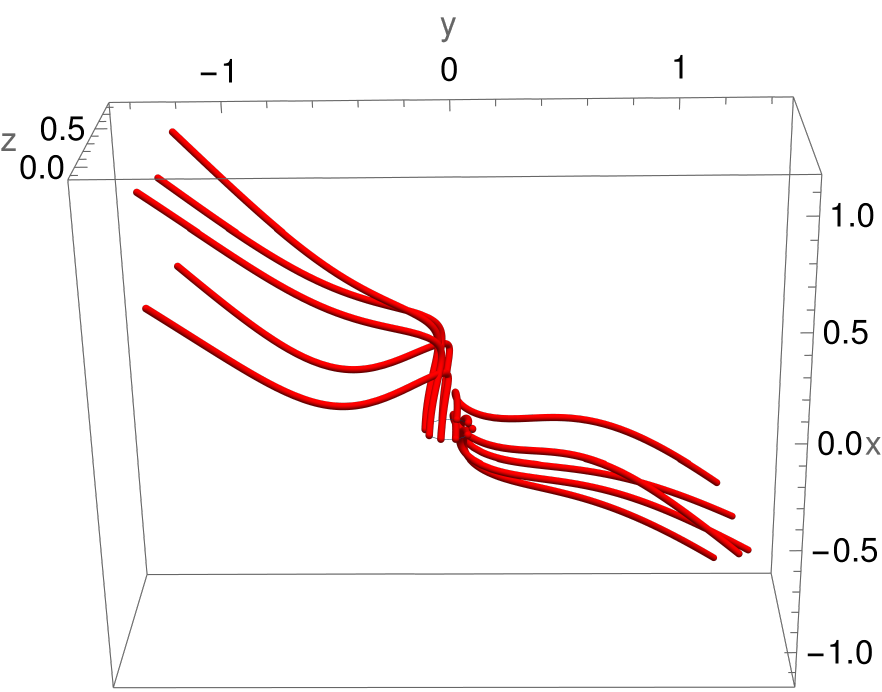}\hspace{1cm}
   \includegraphics[width = 7cm, height = 5cm]{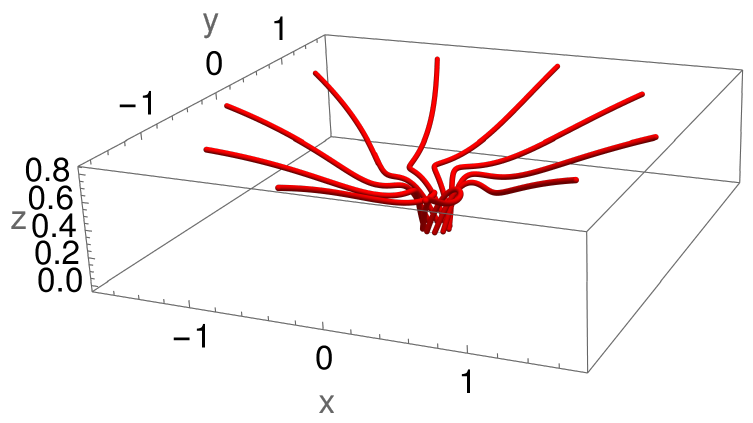}
 \caption{
 Simulation of $N{=}10$ particles in scenario (1B) with $r{=}0.1$ for $(1,0,0)_I$ configuration ($R_{max}=0$) with $\kappa {=} 10$ and $t \in [0,2]$.
 Left: normal direction is $x$-axis. Right: normal direction is $z$-axis.}
\label{2Dpos}
\end{figure}

\begin{figure}[H]
\centering
   \includegraphics[width = 3cm, height = 7cm]{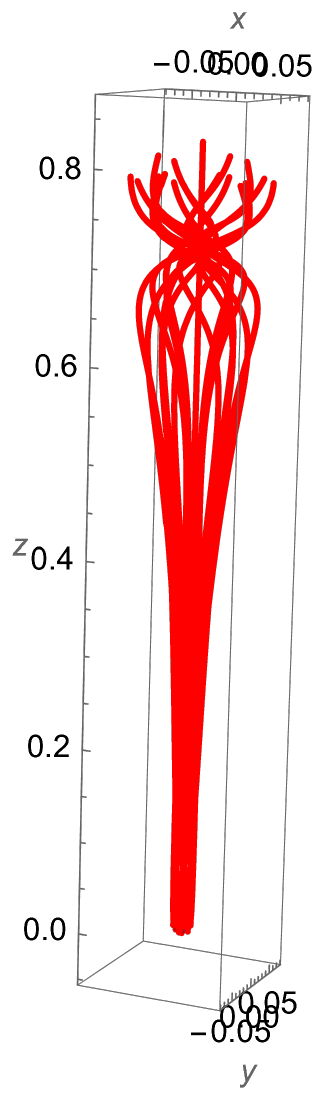}\hspace{2cm}
   \includegraphics[width = 8cm, height = 7cm]{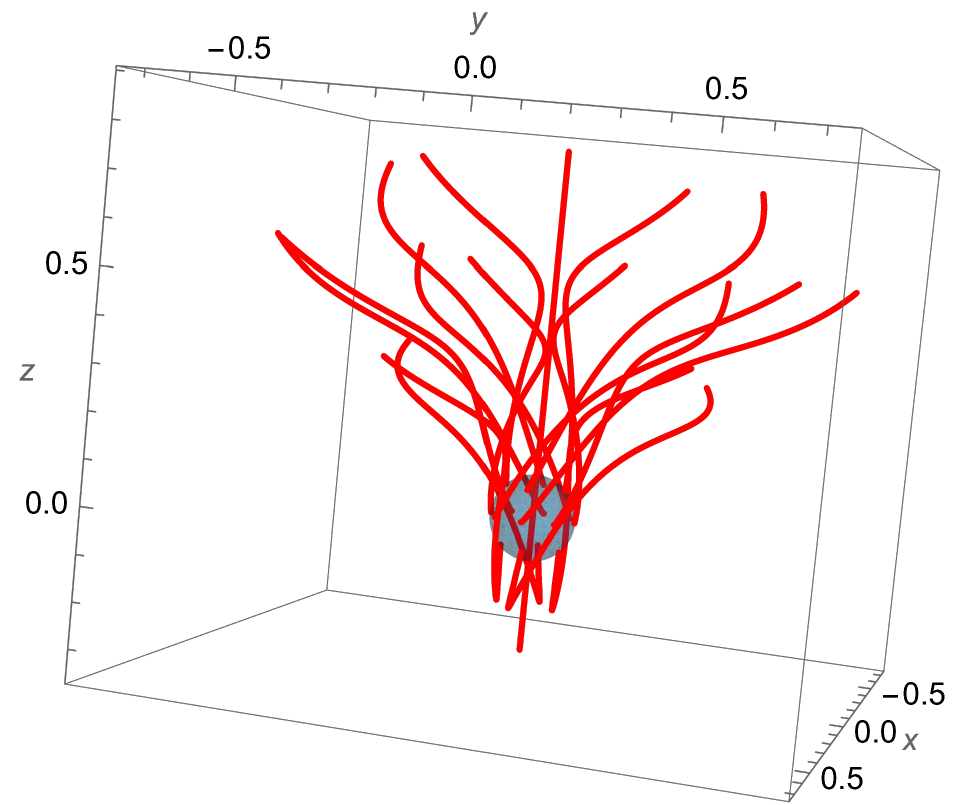}
 \caption{
 Simulation of $N{=}18$ particles in scenario (1C) for $(1,-1,1)_R$ configuration ($R_{\textrm{max}}=0$) with $t \in [0,1]$. Left: $r{=}0.01$ and $\kappa{=}10$. Right: $r{=}0.1$ and $\kappa{=}100$.}
\label{3Dpos}
\end{figure}

We employ the parameter $R_{max}$ \eqref{Rmax} in the the following Figures \ref{1Dpos}, \ref{1Dvel}, \ref{2Dvel}, \ref{3Dvel}, \ref{2Dpos}, and \ref{3Dpos} for both kinds of initial conditions viz.~(1) and (2) (it is especially relevant for the former) to understand the effect of field intensity on particle trajectories.
\begin{figure}[H]
\centering
   \includegraphics[width = 7cm, height = 5cm]{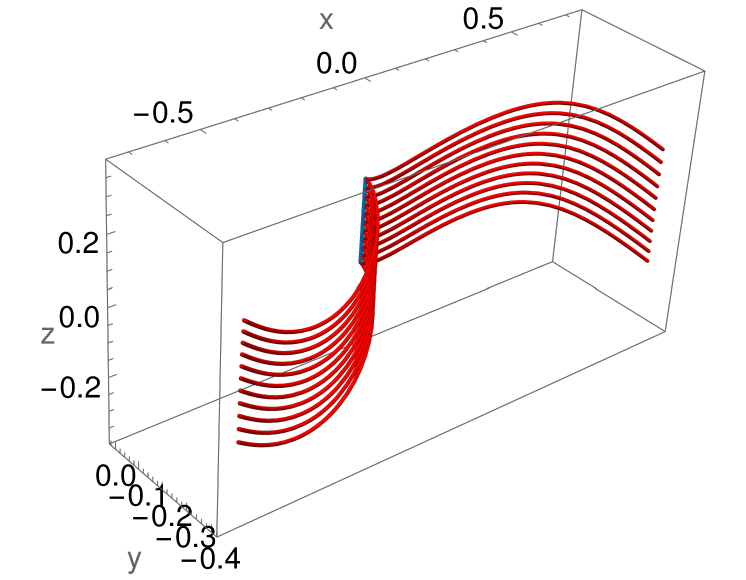}\hspace{1cm}
   \includegraphics[width = 7cm, height = 5cm]{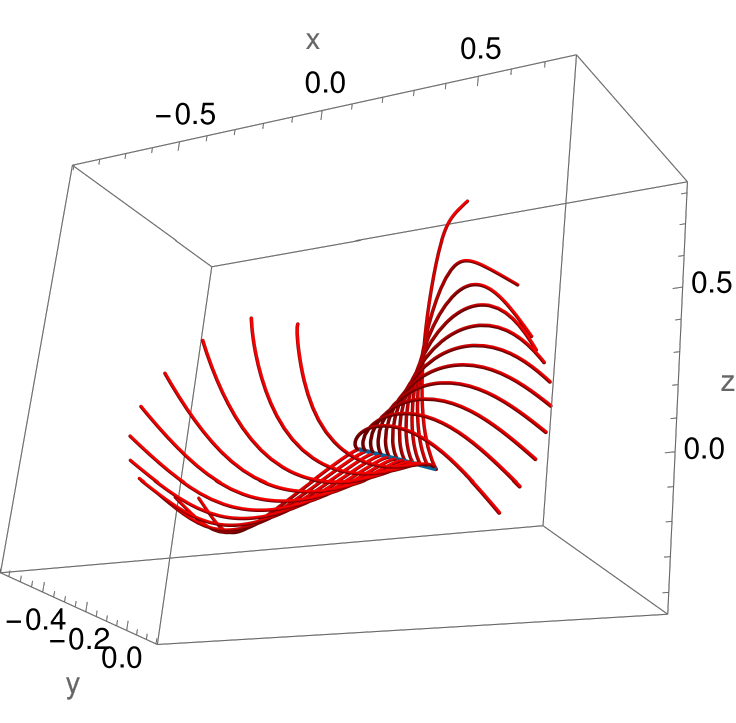}
 \caption{
Simulation of $N{=}11$ particles in scenario (1A) with $|{\bf X}_0| \propto 0.025$ (including one at the origin), for $(\sfrac12;\sfrac12,\sfrac12)_I$ configuration ($R_{max}=0$) with $\kappa{=}10$ and $t\in[-1,1]$. Left: particles initially located along $z$-axis (blue line). Right: particles initially located along some (blue) line in $xy$-plane.}
\label{1Dpos}
\end{figure}

One very interesting feature of trajectories for some of these field configurations is that they twist and turn in a coherent fashion owing to the symmetry of the background field. For particles with initial condition of kind (2), we see that their trajectories take sharp turns, up to two times, with mild twists before going off asymptotically. This has to do with the presence of strong background electromagnetic fields with knotted field lines. This is clearly demonstrated below in Figures \ref{1Dvel}, \ref{2Dvel}, and \ref{3Dvel}. It is worthwhile to notice in Figure \ref{1Dvel} that the particle which was initially at rest moves unperturbed along the $z$-axis; again, this has to do with the fact that these fields have preferred $z$-direction.
\begin{figure}[H]
\centering
   \includegraphics[width = 7cm, height = 5cm]{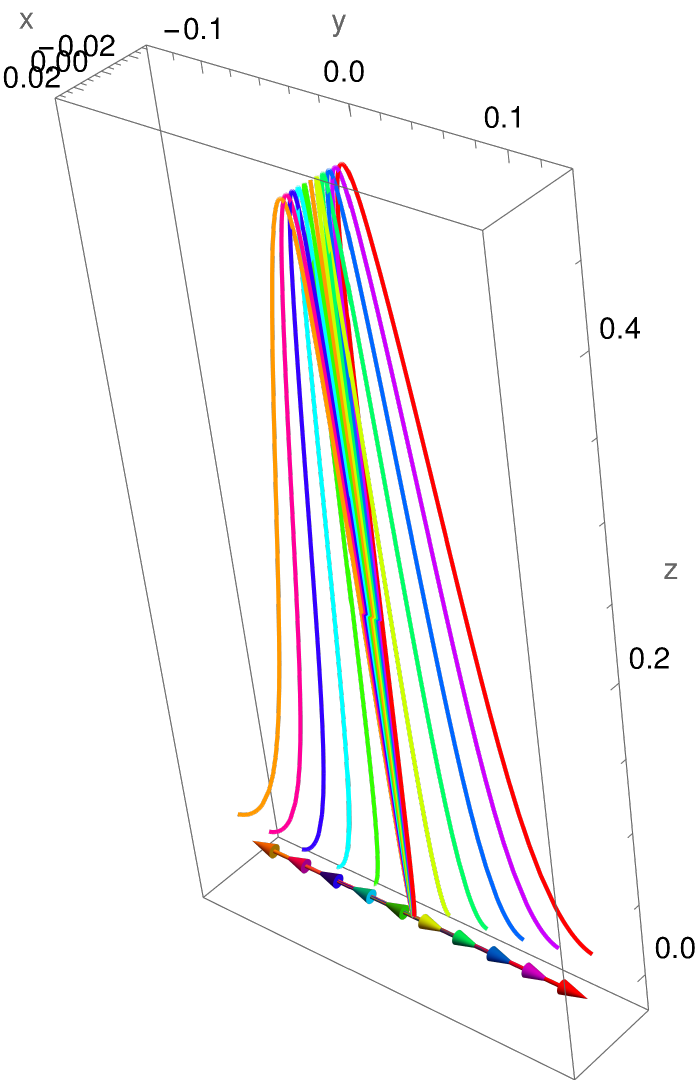}\hspace{1cm}
   \includegraphics[width = 7cm, height = 5cm]{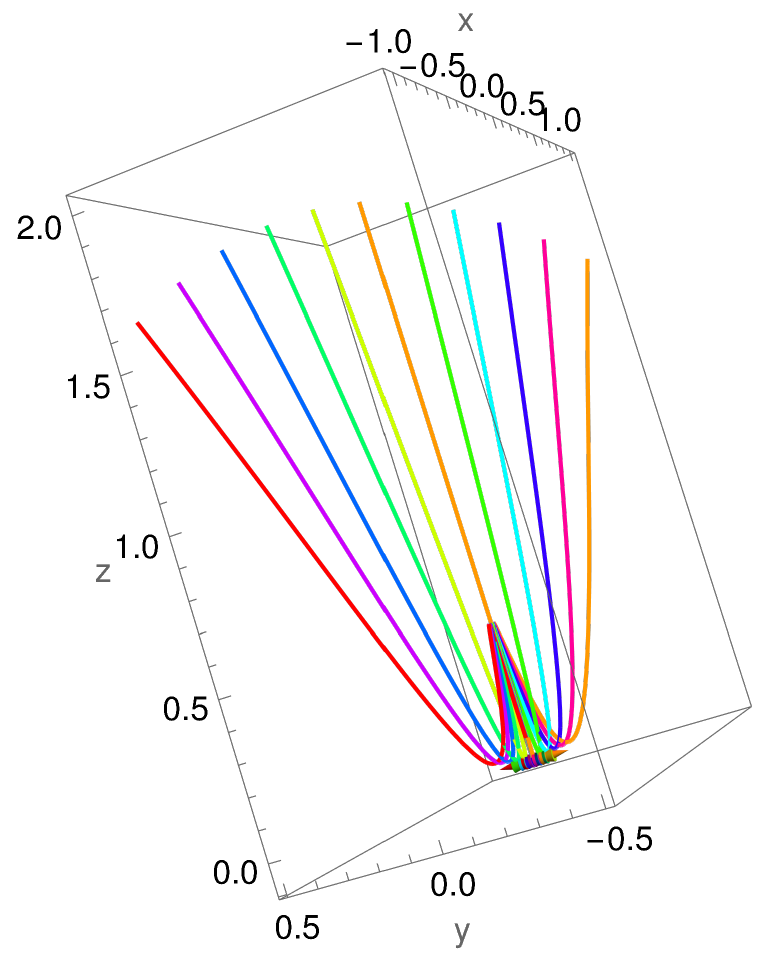}
 \caption{
 Simulation of $N{=}11$ particles in scenario (2A) with $|{\bf V}_0| \propto 0.025$ in the direction of $(0,1,0)$ (including one at rest), for $(1,0,0)_I$ configuration ($R_{max}=0$), with $\kappa{=}10$. Left: $t\in[0,1]$. Right: $t\in[0,3]$.}
\label{1Dvel}
\end{figure}

This feature is even more pronounced in Figure \ref{2Dvel} and (the right subfigure of) Figure \ref{3Dvel} where we see that particles with ultrarelativistic initial speeds are forced to turn (almost vertically upwards) due to the strong electromagnetic field. These particles later take very interesting twists in a coherent manner. This twisting feature is much more refined for the case where initial particle velocities were directed along the $xy$-plane. Here also, we can safely attribute this behavior of the particle trajectories to the special field configurations, with preferred $z$-direction, that we are working with.
\begin{figure}[H]
\centering
   \includegraphics[width = 7cm, height = 5cm]{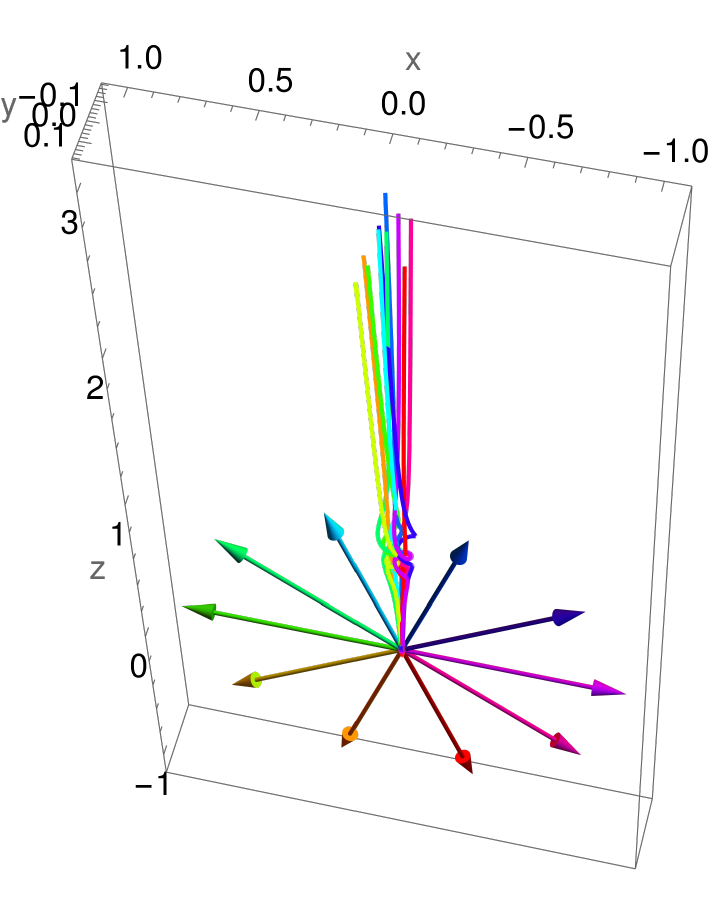}\hspace{1cm}
   \includegraphics[width = 7cm, height = 5cm]{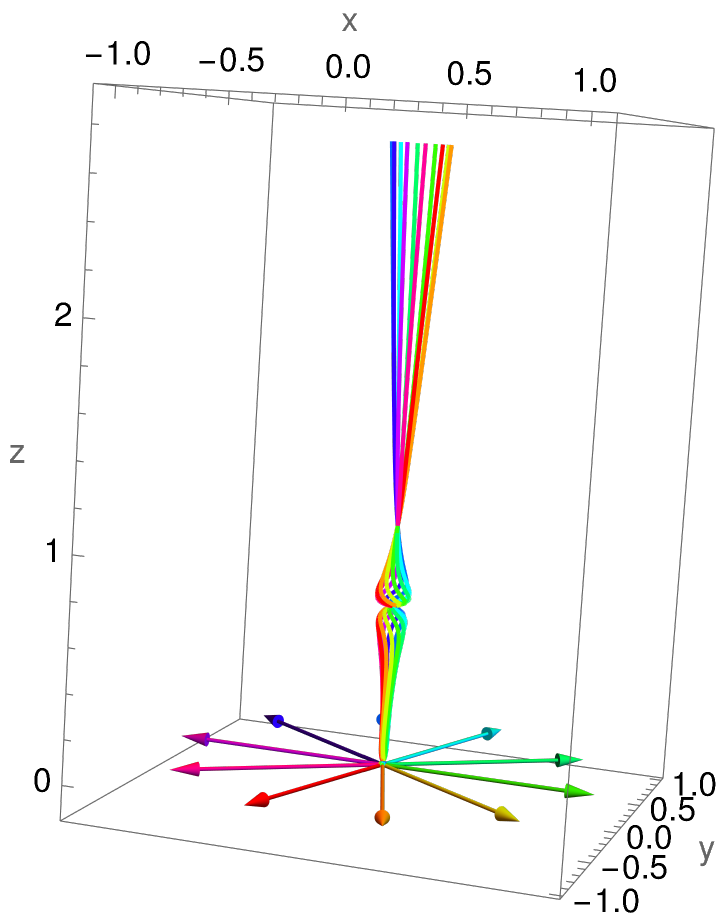}
 \caption{
 Simulation of $N{=}10$ particles in scenario (2B) with $r{=}0.99$ for $(1,-1,1)_R$ configuration with $\kappa{=}100$ and $t \in [0,3]$.
 Left: normal direction is $y$-axis. Right: normal direction is $z$-axis.}
\label{2Dvel}
\end{figure}
\begin{figure}[H]
\centering
   \includegraphics[width = 5cm, height = 7cm]{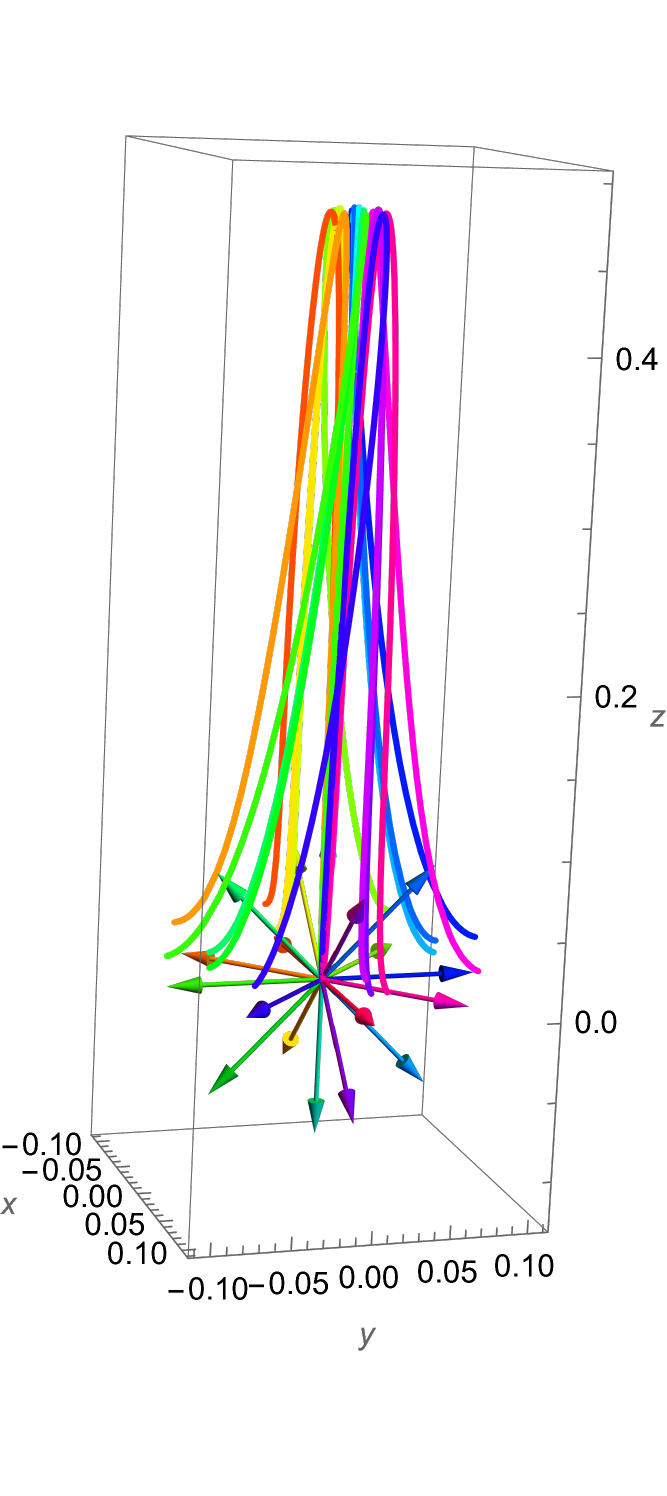}\hspace{2cm}
   \includegraphics[width = 7cm, height = 6.5cm]{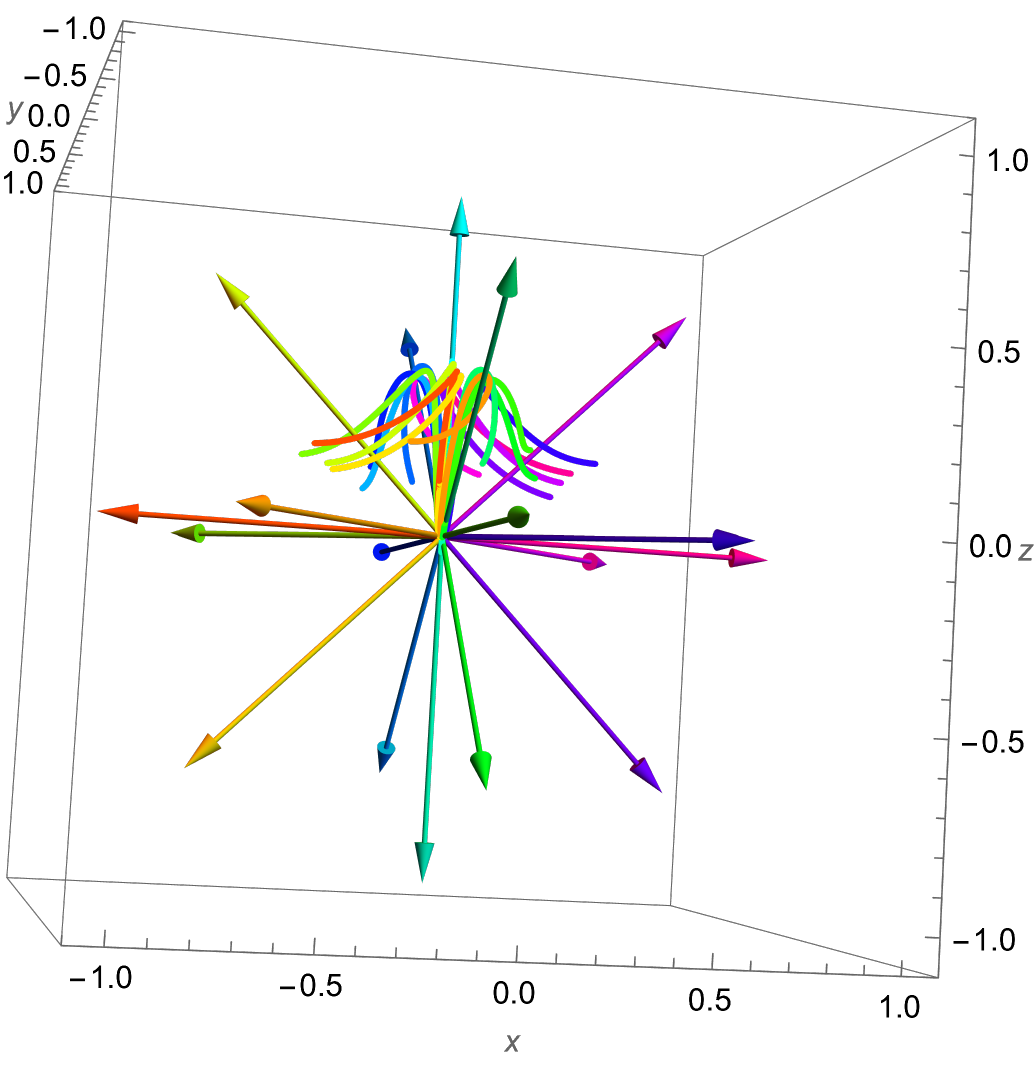}
 \caption{
Simulation of $N{=}18$ particles in scenario (2C) for $(1,0,0)_I$ configuration and $t \in [0,2]$. Left: $r{=}0.1$ and $\kappa{=}10$. Right: $r{=}0.99$ and $\kappa{=}100$.}
\label{3Dvel}
\end{figure}

 We see in Figure \ref{2Dpos} that the trajectories of particles that were initially located on a circle whose normal is along the $z$-axis flow quite smoothly with mild twists for some time before they all turn symmetrically in a coherent way and go off asymptotically. Comparing this with the other case in Figure \ref{2Dpos}, where particles split into two asymptotic beams, we realize that this is yet another instance of the preferred choice of direction for the electromagnetic fields influencing the trajectories of particles.

In Figures \ref{3Dpos} and \ref{2Dvel} we find examples of kind (1) and (2) respectively where both twisting as well as turning of trajectories is prominant. We see in Figure \ref{3Dpos} that the particles that start very close to the origin take a longer time to show twists as compared to the ones that start off on a sphere of radius $R_{max}$. This is due to the fact that the field is maximal at $R_{max}$ and hence its effect on particles is prominent, as discussed before. We also notice here that the particles sitting along the $z$-axis at $T{=}0$ (either on the north pole or on the south pole of this sphere) keep moving along the $z$-axis without any twists or turns. This exemplifies again the fact that these background fields have a preferred direction.

For higher-spin configurations the maximum of the energy density increases but it gets localized into an increasing number of lobes centered around the origin, due to the presence of higher-spin harmonics. Thus, only particles located very close to the tip of these lobes of maximum energy density get accelerated to ultrarelativistic speeds, while particles located outside (which effectively means most of the space) remain unaffected.

\section{Conclusions}

We have discussed the trajectories of charged particles subject to knotted electromagnetic fields generated by the `de Sitter method'. We first reviewed the construction of the fields using the aforementioned method, followed by a discussion of some of their features, including field lines and energy densities in different cases. Afterwards, we discussed trajectories of charged particles in those fields, in different settings.

Various behaviors were obtained by a numerical simulation of the trajectories, including a separation of trajectories into different `solid angle regions' that converge asymptotically into a beam of charged particles along a few particular regions of space, an ultrarelativistic acceleration of particles and coherent twists/turns of the trajectories before they go off asymptotically.

The results contribute to an effort to better understand the interactions between electromagnetic knots and charged particles. This becomes increasingly relevant as laboratory generation of knotted fields progresses. We plan to comprehensively study how exactly the family of torus knots, obtained from a Seifert fibration or via Bateman's construction, is related to our basis configurations. Another future work in this direction may be to analyze a single Fourier mode of these solutions to understand its experimental realization via monochromatic laser beams.

\subsection*{Acknowledgments}

K.K. thanks Deutscher Akademischer Austauschdienst (DAAD) for the doctoral research grant $57381412$.

\clearpage


\end{document}